\documentclass[
    reprint,
    prd,
    superscriptaddress,
    amsmath,
    amssymb,
    aps,
    altaffilletter,
    floatfix
]{revtex4-2}
\usepackage{ragged2e}  
\usepackage[dvipsnames]{xcolor}   
\usepackage{graphicx}
\usepackage{dcolumn}
\usepackage{bm}

\usepackage{multirow} 
\usepackage{siunitx}
\usepackage{upgreek}

\long\def\/*#1*/{} 
\usepackage{lineno}
\usepackage[normalem]{ulem}

\usepackage[caption=false]{subfig}

\usepackage{stackengine}
\usepackage{array}
\usepackage{tabularx}
\usepackage{siunitx}
\usepackage[colorlinks=true, allcolors=blue]{hyperref}

\begin{document}

\title{KATRIN Sensitivity to keV Sterile Neutrinos with the TRISTAN Detector Upgrade}

\newcommand{\berlin}{Institut f\"{u}r Physik, Humboldt-Universit\"{a}t zu Berlin, Newtonstr.~~15, 12489 Berlin, Germany}
\newcommand{\bonn}{Helmholtz-Institut f\"{u}r Strahlen- und Kernphysik, Rheinische Friedrich-Wilhelms-Universit\"{a}t Bonn, Nussallee 14-16, 53115 Bonn, Germany}
\newcommand{\chulalongkorn}{Department of Physics, Faculty of Science, Chulalongkorn University, Bangkok 10330, Thailand}
\newcommand{\cmu}{Department of Physics, Carnegie Mellon University, Pittsburgh, PA 15213, USA}
\newcommand{\etp}{Institute of Experimental Particle Physics~(ETP), Karlsruhe Institute of Technology~(KIT), Wolfgang-Gaede-Str.~1, 76131 Karlsruhe, Germany}
%

\newcommand{\iap}{Institute for Astroparticle Physics~(IAP), Karlsruhe Institute of Technology~(KIT), Hermann-von-Helmholtz-Platz 1, 76344 Eggenstein-Leopoldshafen, Germany}
\newcommand{\ipe}{Institute for Data Processing and Electronics~(IPE), Karlsruhe Institute of Technology~(KIT), Hermann-von-Helmholtz-Platz 1, 76344 Eggenstein-Leopoldshafen, Germany}
\newcommand{\itep}{Institute for Technical Physics~(ITEP), Karlsruhe Institute of Technology~(KIT), Hermann-von-Helmholtz-Platz 1, 76344 Eggenstein-Leopoldshafen, Germany}

\newcommand{\infnbicocca}{Istituto Nazionale di Fisica Nucleare (INFN) -- Sezione di Milano-Bicocca, Piazza della Scienza 3, 20126 Milano, Italy}

\newcommand{\infnmilano}{Istituto Nazionale di Fisica Nucleare (INFN) -- Sezione di Milano, Via Celoria 16, 20133 Milano, Italy}

\newcommand{\polimi}{Politecnico di Milano, Dipartimento di Elettronica, Informazione e Bioingegneria, Piazza L. da Vinci 32, 20133 Milano, Italy}

\newcommand{\umilano}{Dipartimento di Fisica, Universit\`{a} di Milano - Bicocca, Piazza della Scienza 3, 20126 Milano, Italy}

\newcommand{\inr}{Institute for Nuclear Research of Russian Academy of Sciences, 60th October Anniversary Prospect 7a, 117312 Moscow, Russia}
\newcommand{\inrfootnote}{Institutional status in the KATRIN Collaboration has been suspended since February 24, 2022}
\newcommand{\lbnl}{Nuclear Science Division, Lawrence Berkeley National Laboratory, Berkeley, CA 94720, USA}
\newcommand{\madrid}{Departamento de Qu\'{i}mica F\'{i}sica Aplicada, Universidad Autonoma de Madrid, Campus de Cantoblanco, 28049 Madrid, Spain}
\newcommand{\mainz}{Institut f\"{u}r Physik, Johannes-Gutenberg-Universit\"{a}t Mainz, 55099 Mainz, Germany}

\newcommand{\mpp}{Max Planck Institute for Physics, Boltzmannstr. 8, 85748 Garching, Germany}

\newcommand{\mpik}{Max-Planck-Institut f\"{u}r Kernphysik, Saupfercheckweg 1, 69117 Heidelberg, Germany}
\newcommand{\massit}{Laboratory for Nuclear Science, Massachusetts Institute of Technology, 77 Massachusetts Ave, Cambridge, MA 02139, USA}

\newcommand{\muenster}{Institute for Nuclear Physics, University of M\"{u}nster, Wilhelm-Klemm-Str.~9, 48149 M\"{u}nster, Germany}
\newcommand{\npi}{Nuclear Physics Institute,  Czech Academy of Sciences, 25068 \v{R}e\v{z}, Czech Republic}
\newcommand{\unc}{Department of Physics and Astronomy, University of North Carolina, Chapel Hill, NC 27599, USA}
\newcommand{\washington}{Center for Experimental Nuclear Physics and Astrophysics, and Dept.~of Physics, University of Washington, Seattle, WA 98195, USA}
\newcommand{\wuppertal}{Department of Physics, Faculty of Mathematics and Natural Sciences, University of Wuppertal, Gau{\ss}str.~20, 42119 Wuppertal, Germany}
\newcommand{\saclay}{IRFU (DPhP \& APC), CEA, Universit\'{e} Paris-Saclay, 91191 Gif-sur-Yvette, France }

\newcommand{\suranaree}{School of Physics and Center of Excellence in High Energy Physics and Astrophysics, Suranaree University of Technology, Nakhon Ratchasima 30000, Thailand}
\newcommand{\tum}{Technical University of Munich, TUM School of Natural Sciences, Physics Department, James-Franck-Stra\ss e 1, 85748 Garching, Germany}
\newcommand{\uhd}{Institute for Theoretical Astrophysics, University of Heidelberg, Albert-Ueberle-Str.~2, 69120 Heidelberg, Germany}
\newcommand{\tunl}{Triangle Universities Nuclear Laboratory, Durham, NC 27708, USA}
%
%
\newcommand{\ornl}{Also affiliated with Oak Ridge National Laboratory, Oak Ridge, TN 37831, USA}
\newcommand{\duke}{Current address: Department of Physics, Duke University, Durham, NC, 27708, USA}
\newcommand{\wnj}{Also affiliated with Department of Physics, Washington and Jefferson College, Washington, PA 15301, USA}
\newcommand{\oxford}{Current address: Department of Physics, University of Oxford, Oxford, UK}
\newcommand{\hll}{Also affiliated with Semiconductor Laboratory of the Max Planck Society, Isarauenweg 1, 85748 Garching, Germany}

\affiliation{\unc}
\affiliation{\tunl}
\affiliation{\iap}
\affiliation{\ipe}
\affiliation{\muenster}
\affiliation{\infnbicocca}
\affiliation{\polimi}
\affiliation{\infnmilano}
\affiliation{\chulalongkorn}
\affiliation{\cmu}
\affiliation{\washington}
\affiliation{\npi}
\affiliation{\etp}
\affiliation{\wuppertal}
\affiliation{\mpik}
\affiliation{\massit}
\affiliation{\tum}
\affiliation{\umilano}
\affiliation{\suranaree}
\affiliation{\duke}
\affiliation{\itep}
\affiliation{\lbnl}
\affiliation{\madrid}
\affiliation{\berlin}
\affiliation{\uhd}
\affiliation{\mainz}

\author{H.~Acharya}\affiliation{\unc}\affiliation{\tunl}
\author{M.~Aker}\affiliation{\iap}
\author{D.~Batzler}\affiliation{\iap}
\author{A.~Beglarian}\affiliation{\ipe}
\author{J.~Beisenk\"{o}tter}\affiliation{\muenster}
\author{M.~Biassoni}\affiliation{\infnbicocca}
\author{B.~Bieringer}\affiliation{\muenster}
\author{Y.~Biondi}\affiliation{\iap}
\author{B.~Bornschein}\affiliation{\iap}
\author{L.~Bornschein}\affiliation{\iap}
% ...

\collaboration{KATRIN Collaboration}

 \date{\today} 

 \begin{abstract}

Sterile neutrinos in the keV mass range are a well-motivated extension of the Standard Model and viable dark matter candidates. Their existence can be probed in laboratory experiments, as the admixture of a sterile state would induce a characteristic kink-like distortion in the $\upbeta$-decay electron energy spectrum.
The KATRIN experiment is designed to measure the effective electron neutrino mass with sub-eV sensitivity by analyzing the endpoint region of the tritium $\upbeta$-decay spectrum. Following the completion of its neutrino mass program, KATRIN will extend its physics reach to the search for keV-scale sterile neutrinos. This effort will be enabled by the TRISTAN detector, a newly developed silicon drift detector array optimized for differential measurements at high rates and energies well below the endpoint.
In this article, we present the projected sensitivity of KATRIN to keV-scale sterile neutrinos using a dedicated simulation framework.
With four months of detector livetime, KATRIN has the statistical power to probe mixing amplitudes at the level of $|U_{e4}|^2 \sim 10^{-6}$ for sterile neutrino masses in the (4--13)\,keV range, significantly extending the reach of previous laboratory searches. The major experimental systematic uncertainties investigated in this work reduces the sensitivity by a factor of 10--50 over the same mass range.

\end{abstract}

 \vspace*{0.4cm}
 \maketitle
 
 \section{Introduction \label{sec:1_Introduction}}

The Karlsruhe Tritium Neutrino (KATRIN) experiment~\cite{KATRIN_LetterIntent_2001, KATRIN_DesignReport_FZKA_2005, KATRIN_DesignConstComm_JINST_2021,  KATRIN_FirstTransmission_JINST_2018, KATRIN_FirstOperation_EPJC_2020}, operated at the Karlsruhe Institute of Technology (KIT), aims to measure the effective mass of the electron antineutrino with sub-eV precision by studying the endpoint region of the tritium $\upbeta$-decay spectrum. 
Since the start of data taking in 2019, KATRIN has successively set the most stringent laboratory limits on the neutrino mass~\cite{KATRIN_NuMass1_PRL_2019, KATRIN_NuMass2_NatPhys_2022, KATRIN_NuMass3_Science_2025}, with the latest result of $m_{\nu} < 0.45\,\mathrm{eV}$ (90\% CL), improving upon the previous direct laboratory bounds~\cite{Mainz_NuMass_EPJC_2005, Troitsk_NuMass_PRD_2011} by about a factor of five. 
Beyond its primary goal, KATRIN has also demonstrated its capability to explore other physics topics by providing constraints on a broad range of new physics scenarios, such as Lorentz invariance violation~\cite{KATRIN_LorentzViolation_PRD_2023}, general neutrino interactions~\cite{KATRIN_GNI_PRL_2025}, and sterile neutrinos in both the eV~\cite{KATRIN_eVSterile_PRL_2021, KATRIN_eVSterile_PRD_2022, KATRIN_eVSterile_Nature_2025} and keV~\cite{KATRIN_keVSterile_EPJC_2023} mass ranges, while also demonstrating sensitivity to light bosons~\cite{KATRIN_LightBosons_Lauer_PoS_2024}.
After the completion of the neutrino mass measurement campaign, KATRIN will extend its reach to keV sterile neutrino searches through integration of the TRISTAN detector~\cite{TRISTAN_Detector_Mertens_JPG_2019, TRISTAN_SDDCharacterization_Mertens_JPG_2020, TRISTAN_ResponseModel_Biassoni_EPJP_2021}. 
This upgrade consists of a novel large-area, highly pixelated silicon drift detector (SDD) array and a dedicated data-acquisition (DAQ) system optimized for high count rates and good energy resolution that will enable KATRIN to perform a differential measurement of the full $\upbeta$-decay spectrum. 
Together with the intense tritium source, the TRISTAN detector will offer sensitivity to a broad range of precision tests in nuclear and particle physics and to searches for physics beyond the Standard Model, such as signals of the emission of new light bosons~\cite{KATRIN_LightBosons_Arcadi_JHEP_2019}, general neutrino interactions~\cite{KATRIN_PhDThesis_Fengler_2025, KATRIN_ExoticCC_Ludl_JHEP_2016}, extra dimensional sterile neutrinos~\cite{SterileSignature_ExtraDim_Rodejohann_PLB_2014}, and right-handed currents~\cite{KATRIN_RHCurrent_Barry_JHEP_2014}. The primary objective of the program is, however, the detection of keV-scale sterile neutrinos, which constitute viable dark-matter candidates~\cite{SterileDM_DodelsonWidrow_PRL_1994}.

The nature of dark matter remains one of the most central open questions in modern physics. Although its existence has been unambiguously established by numerous astrophysical and cosmological observations, 
the Standard Model does not include any particle that can account for the observed dark matter abundance~\cite{DMReview_Cirelli_arXiv_2024}. New physics is therefore required to explain its fundamental nature~\cite{DMReview_Candidates_Feng_ARAA_2010}.

A sterile neutrino with a mass in the keV range represents a well-motivated dark matter candidate, requiring only a minimal extension of the Standard Model~\cite{SterileDM_DodelsonWidrow_PRL_1994, SterileDM_nuMSM_Asaka_PLB_2005a, SterileDM_nuMSM_Asaka_PLB_2005b, SterileDM_WhitePaper_JCAP_2017}. 
Sterile neutrinos are hypothetical neutrino states that do not take part in the weak interaction. In the presence of active-sterile mixing, the neutrino mass eigenstates are admixtures of active interaction states and sterile right-handed states. The heavy state considered here is predominantly sterile, with a small active admixture that allows it to couple weakly to beta decay. It is characterized by its mass $m_4$ and its active-sterile mixing parameter $\sin^2\theta$.
Depending on their production mechanism, keV-scale sterile neutrinos can constitute so-called warm dark matter, which is often discussed in connection with small-scale structure anomalies such as the \textit{missing-satellite problem}~\cite{DMStructure_MissingSatellites_Klypin_ApJ_1999} and the \textit{cusp-core problem}~\cite{DMStructure_CuspCore_Moore_Nature_1994,DMStructure_CuspCore_Flores_ApJL_1994}. In these cases, standard cold-dark-matter simulations predict an excess of small structures compared to observations. Warm dark matter can alleviate these tensions while remaining consistent with the observed large-scale structure of the Universe~\cite{SterileDM_WarmDM_Bode_ApJ_2001}.
The allowed parameter range for keV sterile neutrinos as dark-matter candidates is mainly motivated by cosmological and astrophysical constraints. These constraints are summarized later in Fig.~\ref{fig:keV_limits}:
\begin{itemize}
    \item As fermions, sterile neutrinos must satisfy the Tremaine–Gunn bound, which sets a lower mass limit of $\sim$0.5\,keV to avoid violating the Pauli exclusion principle~\cite{SterileDM_PhaseSpace_TremaineGunn_PRL_1979}. 
    \item Small-scale structure data can set more stringent bounds on the mass. These bounds depend on the specific production mechanism of sterile neutrinos, which determines their temperature profile. While warm dark matter can alleviate tensions of small-scale structures, too warm (or even hot) dark matter candidates are ruled out, as they would wash out the small-scale structure to an extent which becomes incompatible with observations.  As a result, the lower bound inferred from the Tremaine--Gunn constraint can be significantly higher, reaching several keV when cosmological production scenarios are taken into account~\cite{SterileDM_LowerBound_Boyarsky_JCAP_2009}.
    \item A stringent upper bound on the mixing angle arises from sterile neutrino production via mixing with active neutrinos (the freeze-in mechanism). If the mixing angle is too large, the resulting abundance of keV-scale sterile neutrinos would exceed the dark matter density observed in the Universe.
    \item In contrast, mixing angles that are too small would not allow the production of a sufficient abundance of keV-scale sterile neutrino dark matter. Specific production mechanisms, such as the Shi–Fuller mechanism~\cite{SterileDM_ShiFuller_PRL_1999}, can predict the correct relic abundance, with sterile neutrino masses and mixing angles lying within the allowed parameter space. 
    This minimal mixing typically lies in the range $\sin^2(2\theta) \sim (10^{-10}-10^{-7})$ for keV-scale masses from 1 to 50\,keV~\cite{SterileDM_DodelsonWidrow_PRL_1994, SterileDM_ColdDM_Abazajian_PRD_2001}.
    \item Another stringent bound arises from the decay of sterile neutrino dark matter. Although sterile neutrino dark matter can decay via mixing with active neutrinos, the mixing is so small that its lifetime is long enough to account for the dark matter observed today. The decay mode into an active neutrino and a mono-energetic X-ray photon would lead to a characteristic signature for X-ray telescopes. 
    The absence of a significant signal places strong constraints, typically imposing an upper bound of $\sin^2(2\theta)$ well below $10^{-7}$~\cite{SterileDM_Review_Boyarsky_PPNP_2019}, particularly at higher sterile-neutrino masses where the decay rate increases with mass.
\end{itemize}

While such astrophysical observations provide valuable information on keV sterile neutrino dark matter, they rely on  assumptions about the underlying cosmological model, the production mechanism, and the abundance and distribution of the particle in the cosmos. Various studies show that the above-mentioned constraints can be alleviated when extending the model assumptions. For instance, in~\cite{SterileDM_TimeDependentMixing_Goertz_arXiv_2025} the late phase transition of an auxiliary scalar field would turn on the active-sterile mixing only at a very late stage in the evolution of the universe. This scenario would avoid any over-production of sterile neutrino dark matter and alleviate X-ray constraints. In another study~\cite{KATRIN_DMProspects_PRD_2019}, these constraints are circumvented by reducing the sterile neutrino contribution to the total amount of dark matter, by postponing dark matter production to lower temperatures, and by reducing the decay rate in photons and active neutrinos through cancelation with a new physics diagram.

Ultimately, direct laboratory measurements are desired to unambiguously establish the existence of sterile neutrinos, without relying on astrophysical assumptions. A powerful method to probe sterile neutrinos in the laboratory is via kinematic signatures. Such searches can be performed in high-precision studies of nuclear decays, for instance in electron capture~\cite{SterileSearch_HUNTER_QST_2021,SterileSearch_BeEST_PRL_2021} or, as of interest here, $\upbeta$-decay. 
Past $\upbeta$-decay experiments have already set constraints down to $|U_{e4}|^2 \sim (10^{-2}-10^{-3})$ for sterile neutrino masses up to $\sim$$50$\,keV~\cite{SterileSearch_Mainz_EPJC_2013, SterileSearch_Troitsk1_JPG_2013, SterileSearch_Troitsk2_JETPL_2017, SterileSearch_Tritium_Hiddemann_JPG_1995, KATRIN_keVSterile_EPJC_2023}. The next improvement in sensitivity is expected to come from experiments able to collect significantly higher-statistics~\cite{TRISTAN_Concept_Mertens_JCAP_2015}.

In this paper, we present an overview study of the KATRIN sensitivity to keV-scale sterile neutrinos in the upcoming phase with the TRISTAN detector. The paper is organized as follows. In Sec.~\ref{sec:2_Theoritical_framework}, we summarize the physics motivation for keV-scale sterile neutrino searches in tritium $\upbeta$-decay. Sec.~\ref{sec:3_Experimental_setup} provides an overview of the experimental setup, including both the baseline configuration for the neutrino-mass measurement and the upgraded configuration for the keV-sterile neutrino search phase. 
The analysis framework, including the modeling and the treatment of dominant systematic effects, is presented in Sec.~\ref{sec:4_Analysis_framework}.
The resulting sensitivities and the impact of systematic uncertainties are presented and discussed in Section~\ref{sec:5_Results}.
Finally, conclusions are provided in Secs.~\ref{sec:6_Conclusion}.

 \section{Theoretical Framework \label{sec:2_Theoritical_framework}}

The $\upbeta$-decay of molecular tritium,
\begin{equation}
\mathrm{T_2} \;\rightarrow\; {}^3\mathrm{HeT}^+ + e^- + \bar{\nu}_e , 
\end{equation}
is a superallowed transition with an endpoint energy of $E_0 \simeq 18.6$\,keV and a short half-life of 12.3\,years.  
The differential $\upbeta$-decay spectrum can be written as
\begin{align}
\frac{d\Gamma}{dE_e}(m_\nu) \;\propto\;&\; 
F(Z,E_e)\,p_e E_e (E_0 - E_e) \nonumber \\
&\times \sqrt{(E_0 - E_e)^2 - m_\nu^2}\,C(E_e),
\end{align}
where $E_e$ and $p_e$ are the electron energy and momentum, $Z$ is the nuclear charge of the daughter nucleus, $F(Z,E_e)$ is the Fermi function describing the Coulomb interaction with the daughter nucleus, $m_\nu$ is the effective electron-neutrino mass, and $C(E_e)$ accounts for theoretical corrections beyond the Fermi function, including radiative, recoil, atomic, and molecular effects~\cite{BetaSpectrum_WILKINSON_NPA_1991, KATRIN_BetaSpectrum_Kleesiek_EPJC_2019, KATRIN_FSD_Saenz_EPJC_2024}.

\begin{figure}[t]
  \centering
  \includegraphics[width=\columnwidth]{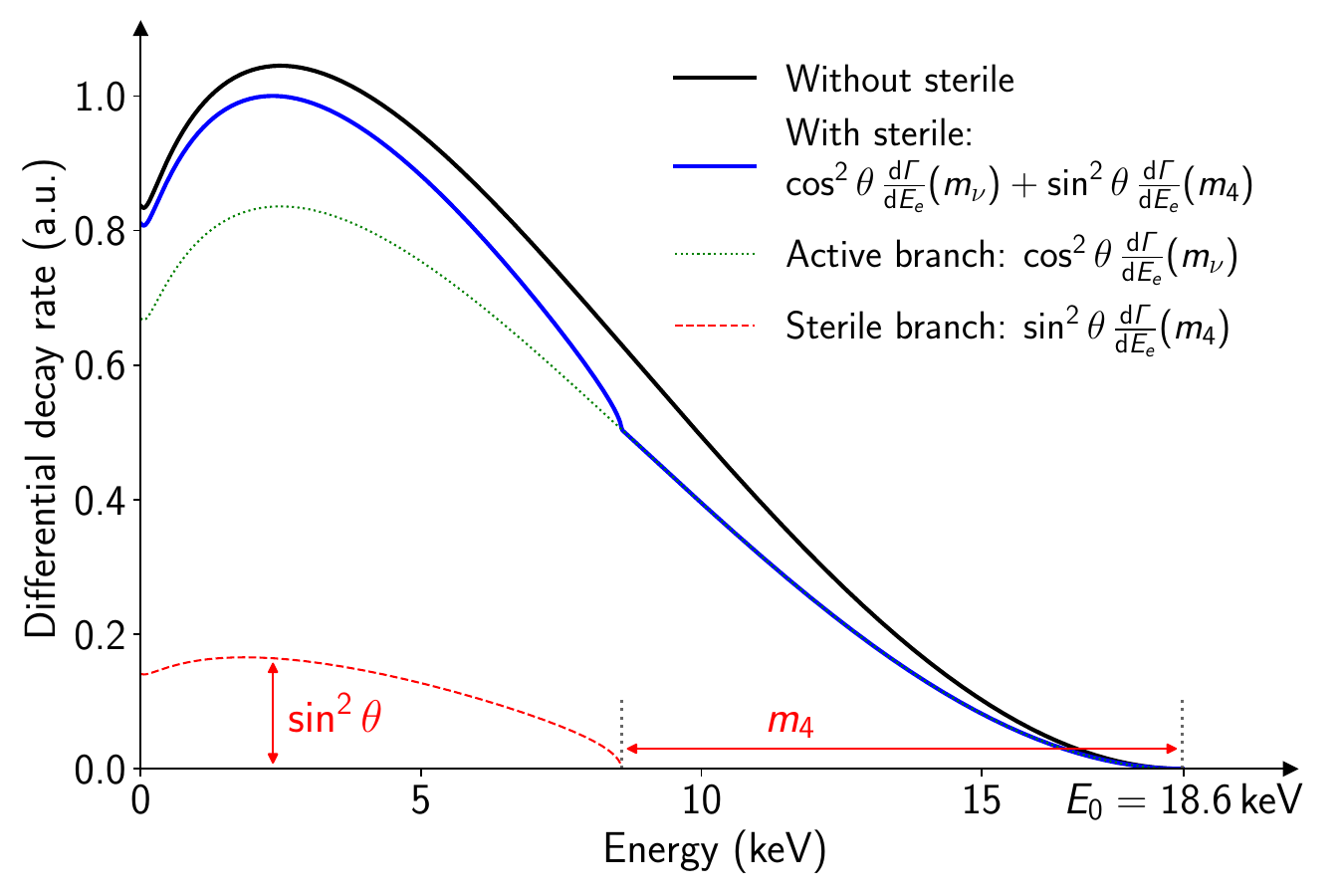}
  \caption{Illustration of a sterile neutrino signature in the theoretical tritium $\upbeta$-decay spectrum. The black curve shows the standard spectrum, while the blue curve includes a sterile state with $m_4=10$\,keV and an unphysically large mixing amplitude of $\sin^2\!\theta=0.2$ for visibility. The kink appears at $E=E_0-m_4$. For clarity, no experimental effects are included in this plot.
}
  \label{fig:kink_spectrum}
\end{figure}

If a sterile neutrino mixes with the electron neutrino, it would produce a distinct signature in the tritium $\upbeta$-decay spectrum, as the decay acquires an additional branch in which the heavier mass eigenstate $m_4$ is emitted~\cite{SterileSearch_NewBounds_Shrock_PLB_1980, SterileSignature_Betadecays_Vega_NPB_2013, SterileSignature_ExtraDim_Rodejohann_PLB_2014}. The resulting differential spectrum is: 
\begin{equation}
\frac{d\Gamma}{dE_e} \;=\; (1-|U_{e4}|^2)\,\frac{d\Gamma}{dE_e}(m_\nu)
+ |U_{e4}|^2\,\frac{d\Gamma}{dE_e}(m_4),
\end{equation}
consisting of the superposition of an active branch governed by the effective neutrino mass $m_\nu$, and a sterile branch determined by $m_4$.
The sterile admixture introduces a kink-like distortion in the spectrum at an energy $E_e = E_0 - m_4$. Its amplitude is proportional to $|U_{e4}|^2 = \sin^2\!\theta$, the active--sterile mixing parameter corresponding to the mixing angle $\theta$ in the two-neutrino framework \footnote{In the two-neutrino mixing scheme, $\sin^2(2\theta) = 4\,|U_{e4}|^2\,(1 - |U_{e4}|^2) \simeq 4\,|U_{e4}|^2$ for small mixing angles.}.
The impact of a sterile neutrino on the tritium $\upbeta$-decay spectrum is illustrated in Fig.~\ref{fig:kink_spectrum}. For tritium, the search is thus kinematically limited to sterile neutrino masses below the endpoint energy, $m_4 \lesssim 18.6$~keV. 

Numerous isotopes with endpoint energies in the few-to-few-tens of keV range, such as $^{3}$H~\cite{SterileSearch_Mainz_EPJC_2013, SterileSearch_Tritium_Hiddemann_JPG_1995, SterileSearch_Troitsk1_JPG_2013, SterileSearch_Troitsk2_JETPL_2017, KATRIN_keVSterile_EPJC_2023}, $^{7}$Be~\cite{SterileSearch_BeEST_PRL_2021}, $^{20}$F~\cite{SterileSearch_20F_Deutsch_NPA_1990}, $^{35}$S~\cite{SterileSearch_35S_Holzschuh_PLB_2000}, $^{63}$Ni~\cite{SterileSearch_63Ni_Holzschuh_PLB_1999}, $^{64}$Cu~\cite{SterileSearch_64Cu_Schreckenbach_PLB_1983}, $^{144}$Pr~\cite{SterileSearch_144Ce_Derbin_JETPL_2018}, $^{187}$Re~\cite{SterileSearch_187Re_Galeazzi_PRL_2001} and $^{241}$Pu~\cite{SterileSearch_MAGNETONU_PRC_2026} have been investigated in searches for keV-scale sterile neutrinos.
Among these isotopes, tritium stands out as an isotope of choice: its endpoint energy provides sensitivity across the relevant mass range; the super-allowed transition enables accurate spectral modeling; and the short half-life permits high-intensity sources for precision measurements. 
Historically, the most stringent laboratory limits on sterile neutrinos in the ($0.01$–$2$)\,keV mass range have been obtained from tritium $\upbeta$-decay experiments, including Mainz~\cite{SterileSearch_Mainz_EPJC_2013}, Troitsk~\cite{SterileSearch_Troitsk1_JPG_2013, SterileSearch_Troitsk2_JETPL_2017}, and the work of Hiddemann et al.~\cite{SterileSearch_Tritium_Hiddemann_JPG_1995}.
Most recently, KATRIN has established the leading laboratory-based constraints in the ($0.01$–$1.0$)\,keV mass range using data from its commissioning campaign~\cite{KATRIN_keVSterile_EPJC_2023}. This result represents an important milestone, demonstrating the feasibility of searching for keV-scale sterile neutrinos with KATRIN. The goal of the present work is to demonstrate that KATRIN, equipped with the TRISTAN detector, has the potential to extend this sensitivity by several orders of magnitude and to higher sterile neutrino masses.

 \section{Experimental setup \label{sec:3_Experimental_setup}}
 
\subsection{KATRIN beamline}

KATRIN consists of a 70-meter-long beamline designed for the direct measurement of the absolute neutrino mass via high-precision $\upbeta$-spectroscopy of tritium $\upbeta$-decay. The beamline is composed of two main sections: the Source and Transport Section (STS) and the tritium-free Spectrometers and Detector Section (SDS)~\cite{KATRIN_DesignReport_FZKA_2005, KATRIN_DesignConstComm_JINST_2021}.

Electrons are produced in the 10-meter-long Windowless Gaseous Tritium Source (WGTS), where highly purified molecular tritium gas is continuously injected and diffuses toward both ends of the tube. The gas is then pumped out and recirculated into the tritium loop system.
The WGTS is designed to provide a high and stable activity of up to $10^{11}$\,Bq. In the current KATRIN configuration, it is operated at $79$\,K.
Electrons are magnetically guided from the source through the transport section to the pre-spectrometer and the main spectrometer (MS), i.e., in the downstream direction toward the detector. 
The transport section reduces the tritium partial pressure by more than 14 orders of magnitude using multiple pumping sections to prevent tritium from entering the spectrometer section.
Both spectrometers function as high-pass filters based on the MAC-E filter principle~\cite{MACE_Lobashev_NIMA_1985, MACE_Picard_NIMB_1992}, which combines magnetic adiabatic collimation with an electrostatic retarding potential. They transmit only electrons whose longitudinal energy, i.e., the kinetic energy associated with the momentum component parallel to the magnetic-field direction, exceeds the applied potential.

Electrons are guided through the 23-m-long, 10-m-wide main spectrometer in a gradually decreasing magnetic field under adiabatic conditions. Because the magnetic field varies slowly compared to the electron cyclotron motion, the magnetic moment is approximately conserved along the trajectory.
During this transport, their pitch angle, defined as the angle between the electron momentum and the local magnetic field direction, governs the distribution of kinetic energy between longitudinal and transverse components. Due to conservation of the magnetic moment, most of the transverse energy is converted into longitudinal energy by the time the electrons reach the analyzing plane, leaving only a small residual transverse component. This adiabatic collimation ensures a large angular acceptance with a maximum accepted pitch angle of $51^\circ$ for electrons emitted in the WGTS ($B_{\text{src}} = 2.52$\,T)~\cite{KATRIN_NuMass3_Science_2025}. 
If this condition were to be violated, non-adiabatic transport would induce uncontrolled pitch-angle changes, leading to unwanted reflections and, in some cases, trapping of electrons.
The magnetic field profile along the main spectrometer is shaped by superconducting solenoids and by an external air-coil system, which allows fine adjustment of the field configuration.
The large field ratio between the field at the exit of the spectrometer and the analyzing plane ($B_{\text{max}} = 4.2$\,T, $B_{\text{ana}} = 0.63$\,mT) results in a maximum residual transverse energy of only $\Delta E \approx 2.8~\mathrm{eV}$ at the endpoint energy~\cite{KATRIN_NuMass3_Science_2025}.

Electrons transmitted through the main spectrometer are then accelerated by 10\,kV using a post-acceleration electrode (PAE) and counted at the focal plane detector (FPD), a 148-segment silicon PIN diode array~\cite{KATRIN_FPD_NIMA_2015}, characterized by an intrinsic energy resolution of order 1.5\,keV (FWHM) for 20\,keV incident electrons. However, in the neutrino-mass measurement, the energy resolution is determined by the MAC-E filter, so that the detector resolution is not a limiting factor.
The integral spectrum of the endpoint region is then obtained by recording the detector count rate at successive retarding potentials ($U_\text{ms}$) in a narrow energy window starting a few tens of eV below the endpoint $E_0$, 
where each measurement gives the number of electrons transmitted above the applied threshold potential.

At the opposite end of the beamline lies the rear section. It contains a gold-plated rear wall absorbing the non-transmitted electrons and defining the electric potential relative to the high voltage of the main spectrometer. This section is also equipped with an angle-selective source of monoenergetic photoelectrons (electron gun) for calibration measurements~\cite{KATRIN_egun_Schneidewind_arXiv_2026}.

\subsection{Beamline upgrades for keV-neutrino sterile searches \label{sec:3b_BeamlineUpgrade}}

The sterile-neutrino signature, although often referred to as a kink, manifests as a broad distortion over the whole tritium $\upbeta$-decay spectrum. An integral high-resolution MAC-E filter scan is therefore not optimal for this search, since it both integrates over much of the relevant spectral information and reconstructs the spectrum from successive retarding-potential settings. This also makes the measurement more sensitive to time-dependent source variations between scan steps in the high-statistics regime relevant for TRISTAN. 
Sensitivity to keV-scale sterile neutrinos therefore motivates a differential measurement of the full $\upbeta$-decay spectrum, in which a broad energy interval is recorded simultaneously and the electron energy is determined by the detector response rather than by the retarding potential. 
In this mode, the spectrometer retarding potential is lowered such that a large range of electron energies is transmitted, resulting in orders-of-magnitude higher count rates. The main spectrometer is nevertheless retained for technical reasons and to preserve the option of future integral measurements for cross-checks and systematic-effect investigations.

For this new phase, critical modifications of the beamline are required, as illustrated in Fig.~\ref{fig:beamline}, showing the KATRIN beamline with the TRISTAN detector installed. The main modifications include:

\begin{figure*}[t]
  \centering
  {
    \includegraphics[width=0.95\textwidth]{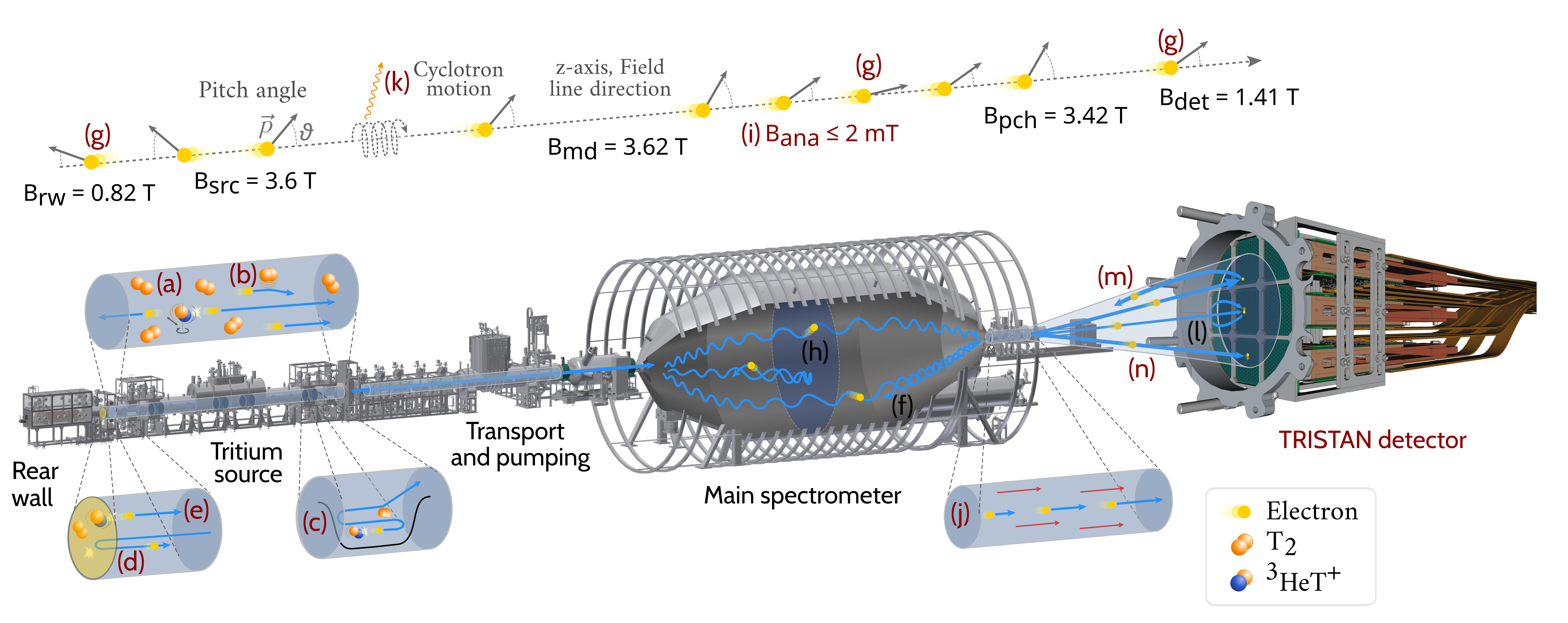}
  }
  \caption{Schematic view of the KATRIN beamline equipped with the TRISTAN detector. The magnetic-field setting corresponds to the reference configuration used in this analysis. The dominant systematic effects relevant for the keV-sterile neutrino search are illustrated: final-state distribution (a), $\mathrm{T_2}$ scattering (b), magnetic traps (c), rear-wall backscattering (d) and decay (e), magnetic reflection (f), magnetic collimation (g), spectrometer transmission (h), spectrometer adiabaticity (i), post-acceleration (j), synchrotron energy loss (k), detector backscattering and back-reflection (l), charge collection and backscattering escape (m), and charge sharing (n).}
  \label{fig:beamline}
\end{figure*}

\textit{Detector system -- }
The most important upgrade is the TRISTAN detection system, a multi-pixel silicon drift detector (SDD) array designed to handle high rates, while replacing the current KATRIN detector section with minimal modifications~\cite{TRISTAN_Detector_Mertens_JPG_2019, TRISTAN_SDDCharacterization_Mertens_JPG_2020, TRISTAN_Detector_Mertens_JPG_2019, TRISTAN_CDR_2022}. 
SDDs are selected for their excellent energy resolution at high count rates, enabled by their small anode capacitance and integrated charge amplification, while the use of drift rings allows for large pixel sizes without compromising performance.
The full TRISTAN detector system will consist of nine modules, each containing a monolithic SDD chip segmented into 166 hexagonal pixels, each with an equivalent radius of about 3\,mm.
The segmentation in modules avoids the technological challenges of producing a single large sensor while ensuring nearly gapless coverage when the modules are placed side by side. The readout electronics are mounted behind the SDD chip, perpendicular to the detector plane, further minimizing insensitive areas~\cite{TRISTAN_ModulesMoS_Siegmann_JPG_2024}. 
In principle, the detector provides an active area of about 12$\times$12\,cm$^2$ with roughly 1500 pixels. 
In practice, only the innermost circular detection area with a diameter of about $9\,\mathrm{cm}$, as defined by the magnetic flux mapping of the source onto the detector, is used.
The TRISTAN modules are designed to be compatible with the strong magnetic fields in the detector chamber and are optimized for a low outgassing rate. Stable operation has been demonstrated in magnetic fields of up to 2\,T, which covers the planned TRISTAN operating conditions.
The prototype and final modules that have already been characterized demonstrate an energy resolution better than 300\,eV FWHM in response to 20\,keV electrons, stable operation under KATRIN-like conditions, and count-rate capabilities up to $\sim$100\,kcps per pixel
~\cite{TRISTAN_ResponseModel_Biassoni_EPJP_2021, TRISTAN_PlanarSetup_CARMINATI_NIMA_2023, TRISTAN_Backscattering_Spreng_JINST_2024, TRISTAN_ModulesMoS_Siegmann_JPG_2024}.
To address the challenges posed by the TRISTAN detector’s high per-pixel count rates and large channel multiplicity, a dedicated DAQ system was designed~\cite{TRISTAN_DAQ_Gavin_arXiv_2026}. This remote analog to digital conversion DAQ relies on early digitization close to the detector and flexible field-programmable gate array (FPGA)-based back-end processing, enabling efficient, high-throughput energy reconstruction with minimal distortion.

\textit{Windowless Gaseous Tritium Source -- }
The WGTS was designed to be operated with a high nominal column density of $\rho d = 5\times 10^{17}$ molecules/cm$^{2}$, where $\rho d$ denotes the integrated number of tritium molecules per unit area along the source axis. For the keV-scale sterile neutrino search, this column density must be adjusted to ensure a detector rate not exceeding $\sim$100\,kcps per pixel. The optimal column density for TRISTAN is projected to be within 0.5–10\% of the nominal value. The exact value will depend on the retarding potential and the magnetic field configuration, both of which also influence the resulting detector rate.

\textit{Post-acceleration electrode -- }
For the neutrino-mass measurement, the PAE is operated at +10 kV (technically limited to +12 kV), primarily to reduce detector-related background by shifting the endpoint region to higher energies.
For this new phase, a PAE capable of sustaining up to +20 kV is under development to reduce detector-related systematic effects, as discussed in Sec.~\ref{sec:4C_Input}.
  
\textit{Rear Wall -- }
Half of the electrons emitted in the WGTS travel upstream and hit the rear wall. 
An additional fraction also reaches this surface after being magnetically reflected in the beamline by the main spectrometer or pinch magnet located at its exit, or after backscattering in the detector.
While many of these electrons are absorbed in the RW, a significant fraction can scatter back into the beamline through various processes. These include single or multiple scattering, as well as the production of secondary electrons via ionization or Auger de-excitation.
The backscattered electrons re-enter the beamline with modified energies and pitch angles and can propagate toward the detector or undergo further reflections. This effect can lead to a significant number of electrons traversing the beamline multiple times before final absorption at the detector. 
While such events do not have a major impact on the neutrino-mass measurement due to the narrow energy window analyzed near the endpoint, they are a significant source of uncertainty for the keV-sterile neutrino search, as most of the scattered electrons will remain within the analysis energy window.
The current RW is gold plated to ensure stable electrical coupling to the source plasma. However, gold (Z=79) exhibits a large probability of electron backscattering due to its high nuclear charge. To mitigate RW backgrounds produced by backscattering, a new RW made of beryllium (Z=4) will be installed. 
Beryllium’s low nuclear charge results in a significantly smaller backscattering coefficient compared to gold (e.g., 3.4\% vs 48\% for 18.6\,keV electrons at normal incidence), with values varying for different incident energies and angles.
 
\begin{figure}[t]
  \centering
  {
    \includegraphics[width=\linewidth]{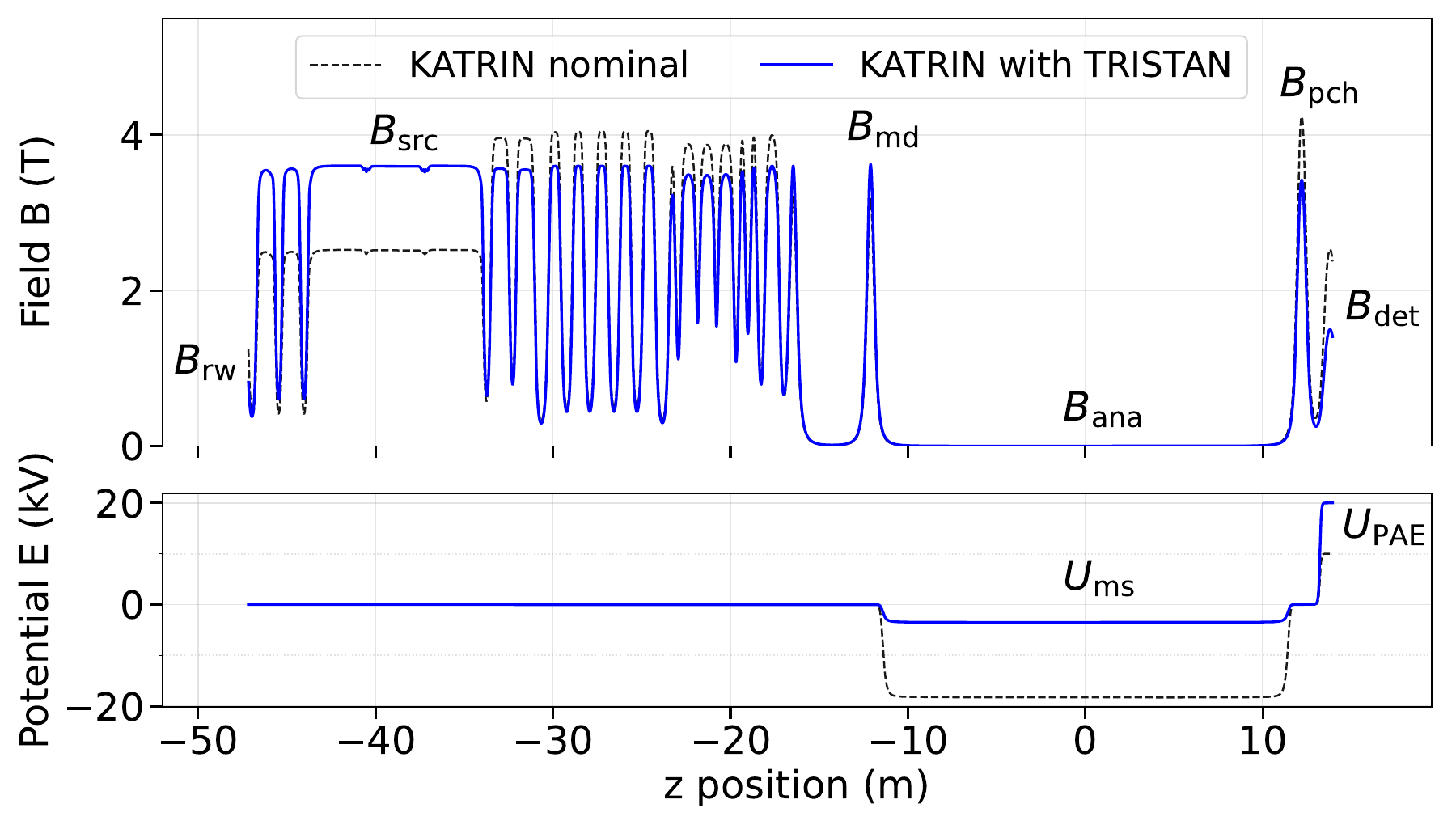}   
  }
  \caption{Comparison of the on-axis magnetic field and electric potential for the nominal KATRIN neutrino measurement campaign and for the keV sterile neutrino search using the TRISTAN detector.}
  \label{fig:global_fields_onaxis}
\end{figure}

\textit{Electromagnetic field design -- }
The magnetic field along the beamline is created by a set of 24 superconducting solenoids delivering maximum fields ranging from 3.6\,T to 6\,T~\cite{KATRIN_Magnets_JINST_2018}. During operation with TRISTAN, the basic principle of electron transport along the beamline remains unchanged, but the main operational parameters are optimized for the keV-sterile neutrino search. 
The magnetic-field and electric-potential configurations assumed for this study are summarized in Table~\ref{tab:beamline_field_parameters} and illustrated in Fig.~\ref{fig:global_fields_onaxis}, where they are compared to the nominal configuration for the neutrino-mass measurement.
\begin{table*}[t]
\centering
\caption{
Summary of the electromagnetic field configuration assumed for this study. The adopted parameters represent a compromise between several competing requirements aimed at mitigating specific systematic effects. Final values will be confirmed with dedicated measurements during commissioning.} 
\begin{tabular}{c l c c}
  \hline
  Parameter & Description & Nominal & TRISTAN \\
  \hline
  $B_\mathrm{rw}$   & RW surface field            & 1.24\,T      & 0.82\,T \\
  $B_\mathrm{src}$  & WGTS field                  & 2.52\,T      & 3.60\,T \\
  $B_\mathrm{md}$   & Max.\ field downstream WGTS & 4.20\,T      & 3.62\,T \\
  $B_\mathrm{ana}$  & MS analyzing-plane field    & 0.6\,mT      & 2\,mT \\
  $B_\mathrm{pch}$  & Pinch-magnet field          & 4.20\,T      & $B_\mathrm{md}-0.2$\,T \\
  $B_\mathrm{det}$  & Detector surface field      & 2.50\,T      & 1.41\,T \\
  $U_\mathrm{ms}$   & MS retarding voltage        & -18.6\,kV    & -3.5\,kV \\
  $U_\mathrm{pae}$  & PAE detector voltage        & +10\,kV     & +20\,kV \\
  \hline
\end{tabular}
\label{tab:beamline_field_parameters}
\end{table*}
These modifications are designed to reduce the contribution of electrons originating from the RW at the detector while ensuring adiabatic transport of the electrons. 
Additionally, the configuration must guarantee that the flux tube mapped onto the detector does not intersect with any beamline elements.

At the RW, a reduced magnetic field $B_{\text{rw}}$ will be used. This setting ensures that backscattered electrons from the RW with large pitch angles are magnetically reflected back toward the RW, providing an additional chance for absorption. 
The relative contribution of RW electrons to the detector spectrum can be further minimized by increasing the fraction of events that propagate directly from the source to the detector by increasing the magnetic field at the source $B_{\text{src}}$.
This adjustment increases the maximum angle of transmission, thereby increasing the acceptance. These modifications are constrained by the maximum technically feasible field at the source of $B_{\text{src}}$ = 3.6\,T and a minimal field at the RW of $B_{\text{rw}}$ = 0.82\,T to avoid flux tube collisions due to the beam tube geometry.

The detector field $B_{\text{det}}$ must be reduced in tandem with the $B_{\text{rw}}$ to ensure the full magnetic mapping of the detector onto the RW. 
While a lower RW field is beneficial for suppressing backscattering contributions, an excessively low detector field would introduce large systematic uncertainties due to an increase of the electron gyroradii and pixel migration from back-reflection at the detector. Dedicated studies have shown that the detector magnetic field should remain above approximately 1\,T~\cite{TRISTAN_PhDThesis_Nava_2025}.

Further optimizations are required to ensure adiabatic transport of high-energy electrons through the MS. In standard KATRIN operation, the maximal downstream magnetic field \(B_{\mathrm{md}}\) is located at the pinch magnet at the downstream end of the MS. For TRISTAN, this magnetic field maximum is shifted to the source side of the MS. This suppresses non-adiabatic pitch-angle changes of electrons with large surplus energy inside the MS. The effectiveness of this mitigation is governed by the magnetic field difference \(\Delta B = B_{\mathrm{md}} - B_{\mathrm{pch}}\). 
In this work, the maximal field downstream of the source $B_{\mathrm{md}}$, is defined with a small offset of 20\,mT relative to $B_{\mathrm{src}}$, resulting in $B_{\text{md}}$ = 3.62\,T. 
This ensures a well-defined global magnetic-field maximum and slightly reduces the maximum accepted pitch angle in the source, to about $86^\circ$, thereby blocking electrons produced very close to $90^\circ$.
A conservative field difference of $\Delta B = 0.2\,\mathrm{T}$ is then adopted, yielding $B_{\text{pch}}$ = 3.42\,T.
To further improve the adiabatic transport through the main spectrometer, the maximum current of the MS air-coil system will be increased, raising the achievable magnetic field inside the spectrometer up to 2\,mT.

Ultimately, the main spectrometer will be operated at a much lower retarding potential, as this setting directly determines the accessible sterile neutrino mass range. The measurement is foreseen at fixed retarding-potential settings rather than in integral mode, but several fixed settings may be used for systematic investigations. 
While the optimal settings must ultimately be determined experimentally, in particular to validate adiabatic transport and to characterize systematic effects, retarding potentials in the range $U_\mathrm{ms} = -8.5$ to $-3.5\,\mathrm{kV}$ are targeted to provide sensitivity to sterile neutrino masses up to $m_4 \approx 10\text{--}15\,\mathrm{keV}$.

 \section{Analysis framework \label{sec:4_Analysis_framework}}

In the keV sterile neutrino search, KATRIN must analyze the full differential tritium $\upbeta$-decay spectrum rather than only the endpoint region. To this end, a dedicated simulation framework has been developed. This section describes the modeling code, the treatment of dominant systematic effects, and the reference settings used for the sensitivity evaluation.

\subsection{Spectrum modeling framework}
\begin{figure}[t]
  \centering
  {
    \includegraphics[width=\linewidth]{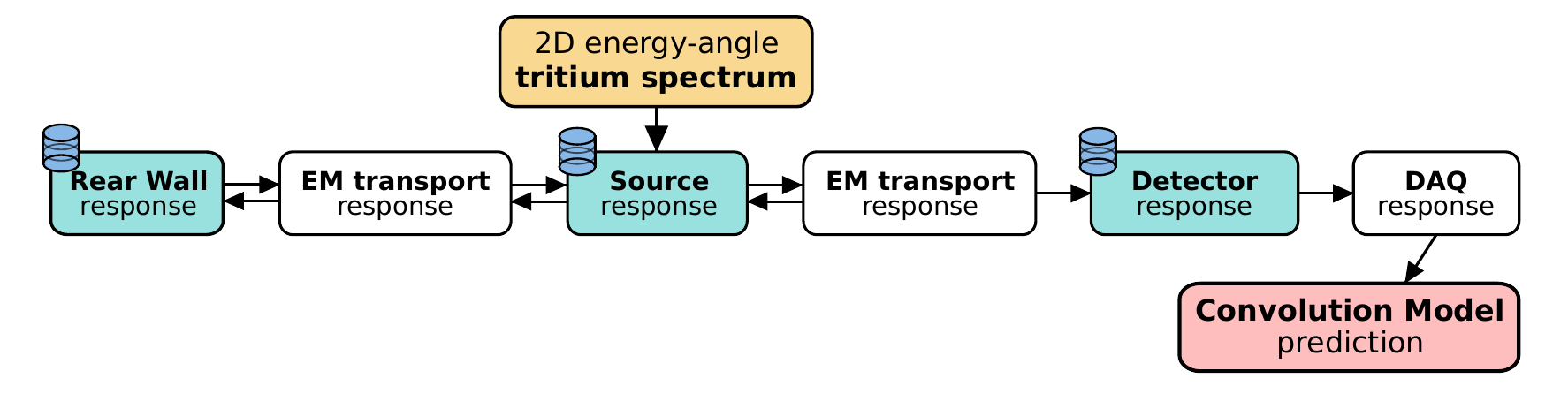}
  }
  \caption{Schematic of the iterative structure of the convolution model used to predict the observable $\upbeta$-decay tritium spectrum. Light-cyan blocks marked with a database symbol denote groups of systematic effects whose responses are implemented using precomputed response matrices stored in dedicated databases.}
  \label{fig:schematic_convmodel}
\end{figure}

The model uses a binned forward convolution method, which consists of folding a theoretical prediction of the differential tritium $\upbeta$-decay spectrum with a series of experimental response functions to predict the measured spectrum at the detector~\cite{TRISTAN_PhDThesis_Descher_2024}. A schematic showing the iterative structure of the code is presented in Fig.~\ref{fig:schematic_convmodel}. 
It is based on a response-matrix approach, in which the input $\upbeta$-decay spectrum is represented as a two-dimensional matrix in the energy–pitch angle phase space, while each experimental effect is described by a four-dimensional matrix that describes how the electron energy and the angle distribution are modified.
After applying a given systematic effect, the output spectrum \(S_{\mathrm{out}}\) is related to the input spectrum \(S_{\mathrm{in}}\) via the integral transform:
\begin{equation}
S_{\mathrm{out}}(E_e',\vartheta') =
\int \!\!\int
R(E_e',\vartheta';E_e,\vartheta)\,
S_{\mathrm{in}}(E_e,\vartheta)\,
\mathrm{d}E_e\,\mathrm{d}\vartheta .
\label{eq:convolution}
\end{equation}
Here \(S(E_e,\vartheta)\) denotes the differential electron rate in energy $E_e$ and pitch angle $\vartheta$, and \(R(E_e',\vartheta';E_e,\vartheta)\) is the response function describing the redistribution of events in \((E_e,\vartheta)\) space.
The transport of  electrons through the beamline is assumed to be adiabatic. It is modeled with an iterative convolution that accounts for changes in energy and direction due to scattering in the source, at the detector or at the RW, as well as electric and magnetic reflections along the beamline.
This model propagates the joint energy–pitch-angle probability density function of the electrons between discrete locations along the beamline, rather than performing full three-dimensional particle tracking in detailed electromagnetic field maps. As a result, it requires only the longitudinal magnetic-field profile and electric potential as shown in Fig.~\ref{fig:global_fields_onaxis}, thereby significantly reducing the computational complexity.
As for other systematic effects, the impact of magnetic and electric fields on the electron energy and pitch angle is encoded in dedicated response matrices. In addition, possible direction reversals due to electric and magnetic mirroring are explicitly taken into account.

The response matrices are generated via dedicated Monte Carlo simulations (custom-developed code or \textsc{Geant4}~\cite{Sim_Geant4_Allison_NIMA_2016}) or analytical calculations and are precompiled into an input library. The matrices are built by scanning the response for all possible electron energies and input angles. The input parameters describing the systematic effects can also be varied in the simulation and embedded as an additional dimension of the response matrix.
The key advantage of this framework is its reliance on pre-generated response matrices, which can be reused once computed without rerunning Monte Carlo simulations. This enables rapid spectrum evaluations, which are essential for studying the impact of systematic uncertainties and for refining or extending individual systematic models without affecting the others. 
In addition, as the response matrices are generated once and reused, the finite Monte Carlo statistics of the underlying simulations are fully correlated across all spectrum predictions. This ensures that differences between spectra arise solely from variations in systematic parameters, rather than from Monte Carlo noise.
The analysis framework is implemented in Python~3 and optimized for performance using vectorized numerical operations, just-in-time compilation, caching, and parallelization. These features allow efficient repeated evaluations required for nuisance-parameter variations and sensitivity studies.
  
\subsection{Systematic effects modeling \label{sec:4B_Syst_modeling}}

The dominant systematic effects accounted for in the model are treated as follows:

\textit{(i) Tritium $\upbeta$-decay and source-related effects --}
The input differential $\upbeta$-decay spectrum is modeled using the standard weak-interaction formalism described in Sec.~\ref{sec:2_Theoritical_framework}, including standard higher-order corrections such as radiative, recoil, screening, and related atomic effects~\cite{KATRIN_BetaSpectrum_Kleesiek_EPJC_2019, BetaSpectrum_WILKINSON_NPA_1991}. Altogether, these corrections modify the spectrum by less than $\sim$1\% above 1\,keV~\cite{TRISTAN_Concept_Mertens_JCAP_2015}.
In $\mathrm{T_2}$ decay, molecular binding energies shift the effective $Q$-value, 
while the sudden change in nuclear charge and molecular configuration leads to excitations described by the final-state distribution (FSD). 
The impact of uncertainties in the theoretical description of the tritium $\upbeta$-spectrum, including the FSD, was previously investigated in a dedicated sensitivity study~\cite{TRISTAN_Concept_Mertens_JCAP_2015}. 
Since then, improved FSD calculations have become available~\cite{KATRIN_FSD_Saenz_EPJC_2024}. Its inclusion in the model with an updated assessment of its impact for the TRISTAN configuration is deferred to future work, as the present study focuses on experimental systematic uncertainties. 

Inside the source, $\upbeta$-decay electrons can scatter off tritium molecules, leading to energy loss and angular changes. This process is modeled using a dedicated KATRIN simulation framework that accounts for electrons produced in the source as well as those re-entering from upstream and downstream. The model incorporates elastic scattering, electronic excitation, and ionization~\cite{Scattering_TotalXS_Liu_PRA_1987,  Scattering_Ionization_Rudd_PRA_1991, Scattering_Elastic_Elinel_JCP_1982, Scattering_Inelastic_Arrighini_MolPhys_1980}, with ionization providing the dominant contribution due to its large cross section and substantial energy loss. 
The scattering probabilities are computed assuming Poisson statistics and are proportional to the product of the energy-dependent cross section and the column density traversed by the electron, where the pitch angle determines the effective path length through the source due to the electron’s cyclotron motion.
Local minima of the magnetic field between the superconducting coils in the source create magnetic traps, in which electrons produced with large pitch angles can oscillate until they escape through scattering. The increased residence time enhances scattering probabilities and modifies both the energy and angle of escaping electrons. 
Magnetic trapping is modeled using a simplified bucket model~\cite{KATRIN_Thesis_Huber_PhD_2021}, with traps approximated as regions of constant magnetic field. The model parameters include the fraction of trapped electrons, estimated from an analytical description of the WGTS longitudinal density profile~\cite{KATRIN_SourceDynamics_Kuckert_Vacuum_2018} and parameterized by an effective  trap column density $\rho d_{\mathrm{trap}}$; and the average magnetic field inside the traps, $B_{\mathrm{trap}}$, derived from simulations with the KASSIOPEIA software package, a dedicated charged-particle tracking framework for simulations in electromagnetic fields developed for the KATRIN experiment~\cite{Kassiopeia_Furse_NJP_2017}.
Finally, energy losses due to synchrotron radiation are neglected, as they have been previously estimated to be negligible compared to the overall energy resolution~\cite{KATRIN_Thesis_Huber_PhD_2021}.

\textit{(ii) Beamline transport and spectrometer effects --}
The model incorporates transport effects for both downstream- and upstream-propagating electrons, assuming adiabatic transport along the beamline such that the orbital magnetic moment of the electron is conserved. 
Under this assumption, transport effects such as magnetic reflection, collimation, spectrometer transmission, and post-acceleration can be treated analytically~\cite{AdiabaticMotion_Tao_PhysPlasmas_2007}. 
Magnetic reflection occurs when electrons propagate from a region of lower to higher magnetic field, leading to a pitch-angle cutoff $\vartheta_{\max} = \arcsin\!\left(\sqrt{B_1/B_2}\right)$ above which electrons are reflected. Conversely, magnetic collimation and de-collimation modify the pitch angle according to the adiabatic relation $\sin^2\vartheta / B = \mathrm{const.}$
Electrons are transmitted through the spectrometer only if their longitudinal kinetic energy exceeds the applied retarding potential. This results in an energy-dependent transmission function whose width is determined by the magnetic-field ratio, $\Delta E \simeq E \cdot B_{\mathrm{ana}} / B_{\mathrm{max}}$. 
For a fixed retarding potential $U_{\mathrm{ms}}$, the transmission width only affects electrons within an energy interval of order $\Delta E$ around the threshold energy $E \simeq qU_{\mathrm{ms}}$. 
For the electromagnetic configuration used here, this corresponds to $\Delta E \simeq 2\,\mathrm{eV}$ at $E \simeq 3.5\,\mathrm{keV}$. 
Electrons with energy $E = qU_{\mathrm{ms}} + \varepsilon$, where $\varepsilon \gg \Delta E$ are fully transmitted.
Downstream of the spectrometer, the PAE increases the electron kinetic energy and decreases the pitch angles at the detector by increasing the momentum component parallel to the magnetic field.
For backward-propagating electrons originating from detector backscattering, reflection at the PAE occurs if their longitudinal kinetic energy is insufficient to overcome the electrostatic potential step.

{\textit{(iii) Rear-wall effects --}
Two distinct effects originate from the RW: (a) partial backscattering and secondary-electron production from incident electrons, and (b) $\upbeta$-decays from tritium accumulated on the rear-wall surface.
Backscattering of electrons on different rear-wall materials is modeled using dedicated \textsc{Geant4} simulations with the PENELOPE low-energy electromagnetic physics package~\cite{Sim_PENELOPE_Sempau_NIMB_1997}. These simulations provide energy- and angle-dependent response functions that map the incident electron distribution at the RW to the spectrum of backscattered and secondary electrons re-entering the beamline.
In addition, tritium accumulation on the RW, originating from the non-negligible residual tritium pressure in the rear-wall region, is modeled as a distinct source term compared to electrons produced in the WGTS. 
As such accumulation is specific to the data-taking conditions and was shown to be subdominant compared to backscattering from electrons produced in the WGTS, this contribution is not considered in the present analysis. Dedicated measurements under representative tritium-exposure conditions are ongoing and will help refine its treatment in future analyses~\cite{TRISTAN_TritiumRW_Batzler_arXiv_2025}.

{\textit{(iv) Detector response --}
The detector response model accounts for three dominant effects: partial energy deposition, intrinsic energy-resolution broadening, and charge sharing between pixels.
Partial energy deposition arises primarily from three processes: 
(a) dead-layer effects, where incomplete charge collection occurs in a thin entrance layer of the detector;
(b) X-ray escape, in which characteristic silicon X-rays ($\sim$1.74\,keV) produced by inelastic interactions exit the sensitive volume; and
(c) backscattering, where scattered electrons leave the detector volume and re-enter the beamline in the upstream direction.
These effects are modeled using a \textsc{Geant4}-based detector simulation combined with an analytic description of the depth-dependent charge-collection efficiency in the detector dead layer. The simulation employs the PENELOPE package, chosen for its demonstrated agreement with dedicated TRISTAN detector-characterization measurements~\cite{TRISTAN_ResponseModel_Biassoni_EPJP_2021, TRISTAN_Backscattering_Spreng_JINST_2024}}. 
The efficiency profile is described by an exponential function characterized by a single length scale parameter $\lambda$. This parameter defines the depth-dependent charge-collection efficiency used to weight individual energy depositions~\cite{TRISTAN_ResponseModel_Biassoni_EPJP_2021}.
The effect of electrons backscattered and backreflected at the detector surface is not explicitly modeled and is instead treated as a systematic uncertainty, as described in Sec.~\ref{subsec:systematics}.
This reflects a limitation of the current framework, as the process depends simultaneously on the detector response and the electromagnetic field configuration. A detailed treatment would necessitate response matrices of substantially higher dimensionality, resulting in prohibitive computational complexity for the present sensitivity study.
Statistical fluctuations in the number of electron--hole pairs generated during energy deposition give rise to Fano noise, resulting in an energy-dependent Gaussian broadening of the spectrum. This contribution is modeled analytically as 
$ \sigma_{\text{fano}}(E) = \sqrt{E \, \varepsilon_{\text{eh}} \, F_{\text{fano}}}$, where $\varepsilon_{\text{eh}} = 3.65\,\mathrm{eV}$ is the mean energy required to create an electron--hole pair in silicon~\cite{DetectorBook_Kolanoski_OUP_2020} and $F_{\text{fano}} = 0.115$ is the Fano factor accounting for the non-Poissonian dispersion of the number of observed charge carriers~\cite{Silicon_Ionization_Alig_PRB_1980}. 
Charge sharing occurs when an electron interacts near a pixel boundary and the resulting charge cloud is collected by adjacent pixels, leading to a reduced energy recorded in the primary pixel and the remaining charge being collected in neighboring channels.
Charge sharing is modeled analytically assuming a Gaussian charge-cloud distribution. The current implementation includes only two-pixel sharing and neglects triple charge sharing, in which the signal is distributed across three pixels. About \( \mathcal{O}(1\%) \) of all events are expected to involve charge sharing, and roughly 5\% of those are expected to be triple charge-sharing events. The model depends on the pixel geometry and on the charge-cloud width $w_{\mathrm{cc}}$.
Triple charge sharing and possible modifications of charge drift and sharing induced by the magnetic field are the subject of ongoing studies. Although charge sharing can be experimentally mitigated using pixel-multiplicity cuts based on temporal coincidences, such cuts are not used in this analysis.

{\textit{(v) DAQ and readout effects --}
The DAQ digitizes detector signals and reconstructs event energies and time stamps through the application of finite-impulse-response filters. These include electronic noise, finite detection threshold, unresolved pileup, deadtime, and energy-scale calibration. Accurate modeling of these effects is critical for high-rate measurements such as TRISTAN where overlapping signals, finite time resolution, and other readout effects can distort the measured spectrum.  In the analysis framework, DAQ-related effects relevant for high-rate operation are modeled at the spectral level. Complementary waveform-level modeling is under development and will be used to further validate and refine this treatment~\cite{TRISTAN_PhDThesis_Urban_2024}.
Electronic noise, originating from intrinsic fluctuations in the readout electronics, is modeled as an energy-independent Gaussian broadening with width $\sigma_{\text{en}}$, added in quadrature to the intrinsic detector resolution. 
A finite detection threshold arises from baseline noise and trigger-filter limitations in the front-end electronics, imposing a minimum signal amplitude required for reliable event detection. This effect is modeled by appling a smooth turn-on at low energies using the cumulative distribution function of a Gaussian with a parametrized threshold position and effective width. 
At the high count rates expected for TRISTAN, multiple events may occur within the minimum pulse-pair resolution time $t_{\mathrm{holdoff}}$, defined by the trigger algorithm. 
Events occurring within this interval cannot be resolved individually and give rise to an unresolved pileup component, in which the deposited energies are summed. This produces a characteristic spectral contribution extending beyond the tritium endpoint. The pileup fraction increases with both the detector rate and $t_{\mathrm{holdoff}}$. 
Unresolved pileup is modeled by self-convolution of the spectrum, corresponding to the random superposition of independent energy depositions within the resolution window. 
The pileup spectrum is constructed before applying the trigger threshold, so the threshold acts only on the total spectrum, without modifying the modeled pileup spectral shape.
The resulting pileup contribution is scaled according to the probability of having two or more events within $t_{\mathrm{holdoff}}$, with higher-order contributions included iteratively up to third order in this analysis.
To prevent overlapping pulses from being processed as a single event, a fixed deadtime window $t_{\mathrm{pu}}$ is applied after each trigger. Events occurring within this deadtime window are rejected and therefore do not contribute to the reconstructed spectrum. Deadtime effects therefore reduce the overall event rate without modifying the spectral shape.
Finally, the conversion from reconstructed pulse heights to physical energy is modeled using a linear calibration characterized by gain and offset parameters, $G$ and $E_{\mathrm{offset}}$ respectively.

{\textit{(vi) Background --}
A small background contribution, originating from a combination of spectrometer-related and detector-related sources such as intrinsic radioactivity and cosmic-muon-induced events, is expected. Detailed measurements of the differential background spectrum at KATRIN have been reported in Ref.~\cite{KATRIN_PhDThesis_Harms_2015}, yielding a rate of approximately $6\times10^{-4}\,\mathrm{cps/keV}$, corresponding to a total background rate of about $10^{-2}\,\mathrm{cps}$ over the measured energy range. For the keV sterile neutrino search, the background level may differ slightly from these measurements due to modified electromagnetic field settings and vacuum conditions associated with the detector upgrade but due to the very high tritium event rate expected, the background contribution will remain negligible compared to the signal. Its impact was therefore investigated by introducing a flat component with a conservative rate of $R_{\text{bkg}} = 1\times10^{-3}\,\mathrm{cps/keV}$.

\subsection{Inputs and reference configuration \label{sec:4C_Input}}

Following the prescription described in Sec.~\ref{sec:3b_BeamlineUpgrade} the set of reference input parameters and uncertainties reported in Table~\ref{tab:systematic_parameters} were used for this analysis. 

In summary, the new beryllium RW was used, and the magnetic field values were optimized to further reduce the contribution of electrons backscattered at the RW.
The detector parameter governing the dead-layer width was set to $\lambda = (58 \pm 2)\,\mathrm{nm}$. 
This choice is consistent with several measurements performed with TRISTAN detector modules and electron-gun data over the years, which generally indicate an effective dead-layer thickness of about 
(50$-$60)\,nm~\cite{TRISTAN_ResponseModel_Biassoni_EPJP_2021, TRISTAN_SDDCharacterization_Mertens_JPG_2020, TRISTAN_PhDThesis_Siegmann_2024}.
The charge-sharing cloud width was set to \(w_{\mathrm{cc}} = (20 \pm 1)\,\mu\mathrm{m}\), guided by dedicated detector characterization studies using electrons, X-rays, and a collimated monochromatic laser beam~\cite{TRISTAN_MScThesis_Urban_2019,TRISTAN_ChargeCollection_Forstner_JINST_2025}. These measurements show that a Gaussian charge-cloud model provides a good description of the data and support a characteristic width in the range of about \(16\text{--}20\,\mu\mathrm{m}\).
Pileup was modeled using a minimum pulse-pair resolution time of $t_{\mathrm{holdoff}} = 112\,\mathrm{ns}$, and a trigger deadtime of $t_{\mathrm{pu}} = 1.15\,\mu\mathrm{s}$. In addition to the intrinsic Fano contribution, an energy-independent Gaussian broadening due to electronic noise with $\sigma_{\mathrm{en}} = 44\,\mathrm{eV}$ was included. More details on the DAQ concept for TRISTAN can be found in~\cite{TRISTAN_DAQ_Gavin_arXiv_2026}.}
The PAE was set to 20\,kV and the retarding potential was set to \(U_{\text{ms}} = -3.5\,\text{kV}\). The energy of the electrons included in the fit therefore span from \(23.5\,\text{keV}\) up to the tritium endpoint at approximately \(38.6\,\text{keV}\), corresponding to an accessible sterile neutrino mass range of up to $\sim$15\,keV.
The column density was set to 0.6\% of the nominal activity to ensure an input electron rate of 100\,kcps per pixel at the detector.  
At this rate, approximately $2\%$ of detected triggers correspond to unresolved pileup, while detector deadtime reduces the effective
livetime to approximately $80\%$. Above this rate, pileup becomes a primary concern by significantly reducing the measured rate. In addition, a rate-independent livetime reduction due to periodic preamplifier reset inhibit windows was included, resulting in an additional $\sim5\%$ loss of livetime for the parameters considered.

Based on magnetic flux–tube mapping at the detector, 524 “golden” pixels, defined as those fully illuminated together with their neighboring pixels, were selected for the analysis. This number reflects the achievable configuration with the current detector chamber design, while further design developments and investigations are ongoing to increase the number of mappable pixels.
With this configuration, a measured rate of 
$\sim$$5\times10^{7}\,\mathrm{e^-/s}$
is expected. 
This rate is several orders of magnitude higher than the average count rate of 
$\sim$$2.4\,\mathrm{e^-/s}$ measured by KATRIN with the FPD during the first five neutrino-mass measurement campaigns for scan steps above $(E_0 - 40)\,\mathrm{eV}$~\cite{KATRIN_NuMass3_Science_2025}.

In the following, unless stated otherwise, a reference data-taking period of four months of detector operation is assumed, corresponding to a total collected statistics of approximately $4\times10^{14}$ recorded events. The resulting expected tritium $\upbeta$-decay spectrum measured with the TRISTAN detector is illustrated in Fig.~\ref{fig:TRISTAN_tritiumspectrum}. 
\begin{figure}[t]
  \centering
  \includegraphics[width=\columnwidth]{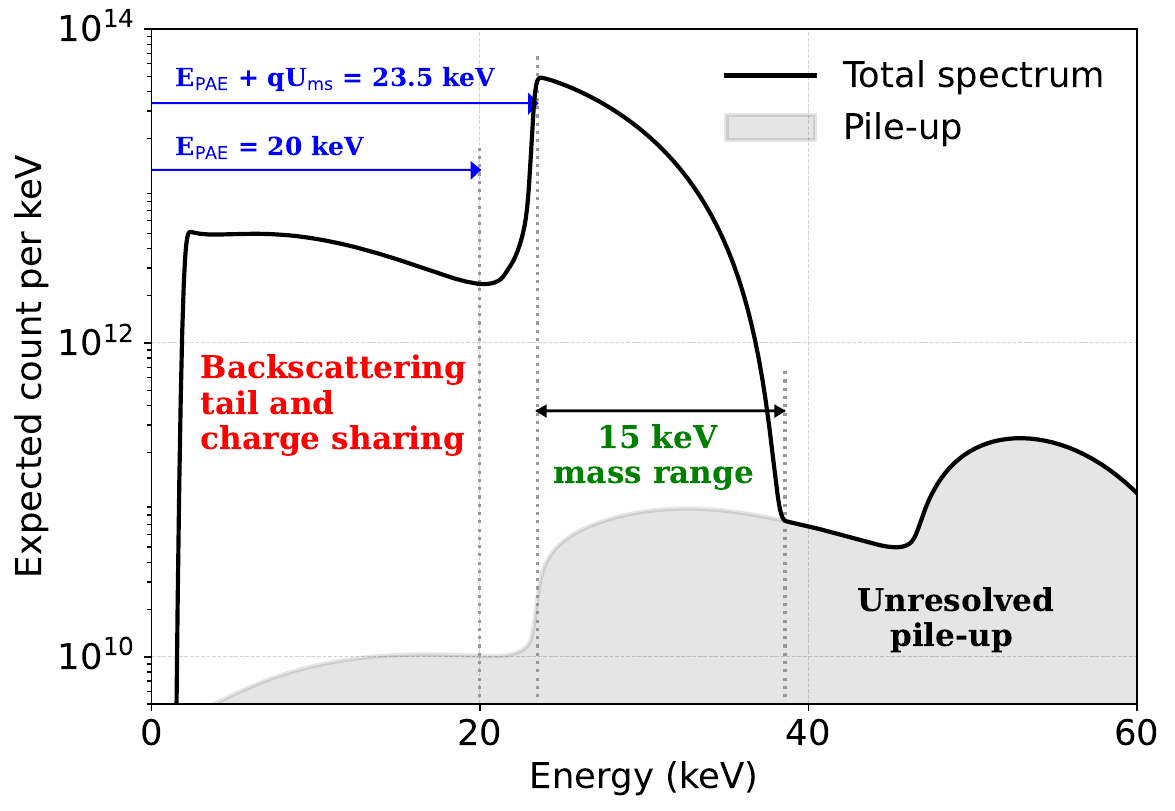}
  \caption{Predicted tritium $\upbeta$-decay spectrum measured by the TRISTAN detector at KATRIN for four months of data taking, in the absence of any keV sterile-neutrino spectral distortion.
  The tritium $\upbeta$-decay spectrum is shifted by 20\,keV due to the acceleration of electrons by the PAE before the detector. Partial energy depositions from charge sharing and backscattering are responsible for the low energy contribution below 20\,keV.
  The non-zero contribution above the tritium energy endpoint is due to unresolved pileup events. In this configuration, because of the $-3.5$\,kV retarding potential, the accessible sterile neutrino mass range is reduced to 15\,keV.
}
  \label{fig:TRISTAN_tritiumspectrum}
\end{figure}
Several effects are visible in the spectrum. The tritium $\upbeta$-decay spectrum is shifted due to the post-acceleration potential, while its low-energy part is truncated by the retarding potential. A low-energy tail induced by detector backscattering and charge-sharing effects, which lead to partial energy deposition, is also visible, together with a non-zero contribution above the tritium endpoint due to pileup events. The mitigation of these effects through the increase of the electron energy before reaching the detector by raising the post-acceleration potential is illustrated in Fig.~\ref{fig:systematic_effects_spectrum}. 
\begin{figure}[t]
  \centering

  \includegraphics[width=\columnwidth]{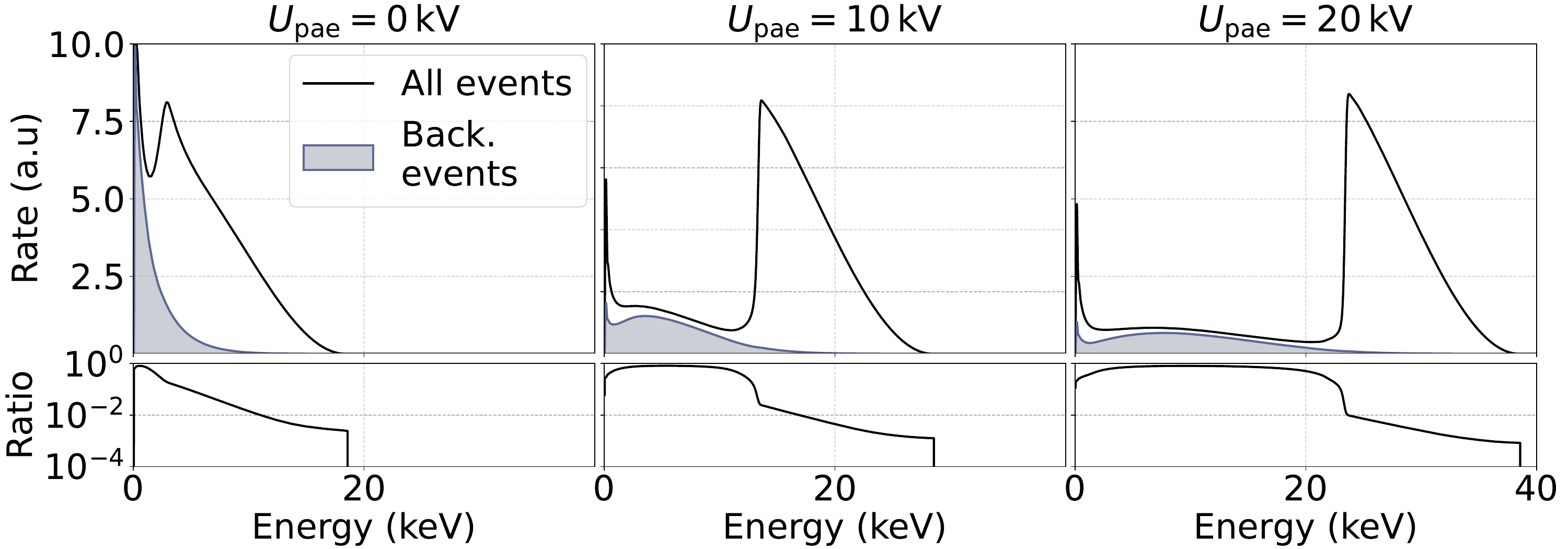}

  {\small (a) Backscattering at the detector (pileup off)\par} 

  \vspace{0.8em}
 
  \includegraphics[width=\columnwidth]{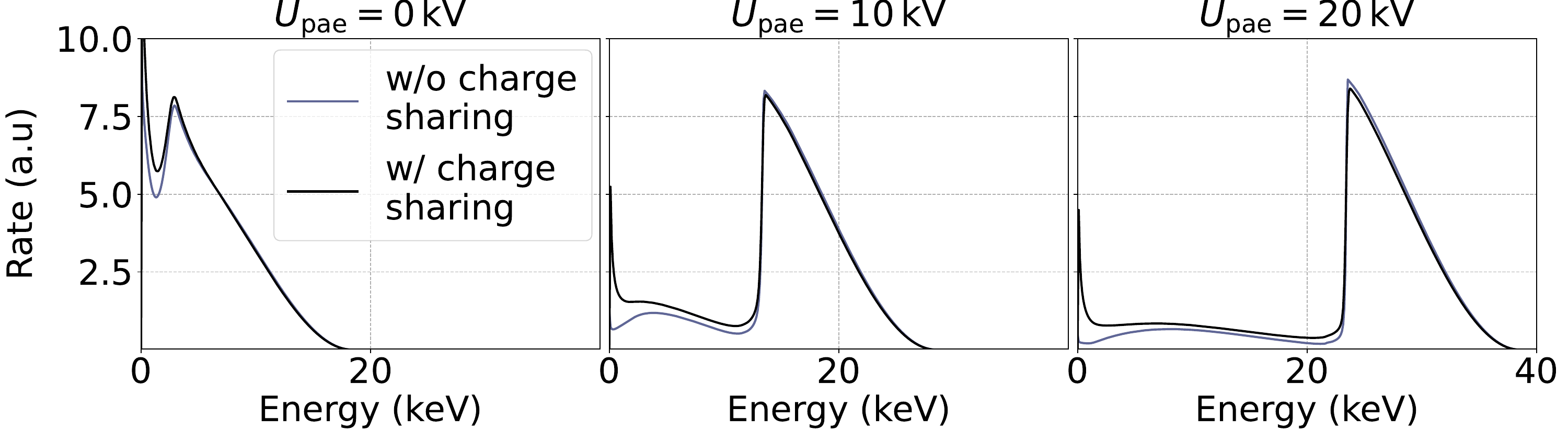}

  {\small (b) Charge sharing (pileup off)\par}

  \vspace{0.8em}
  
  \includegraphics[width=\columnwidth]{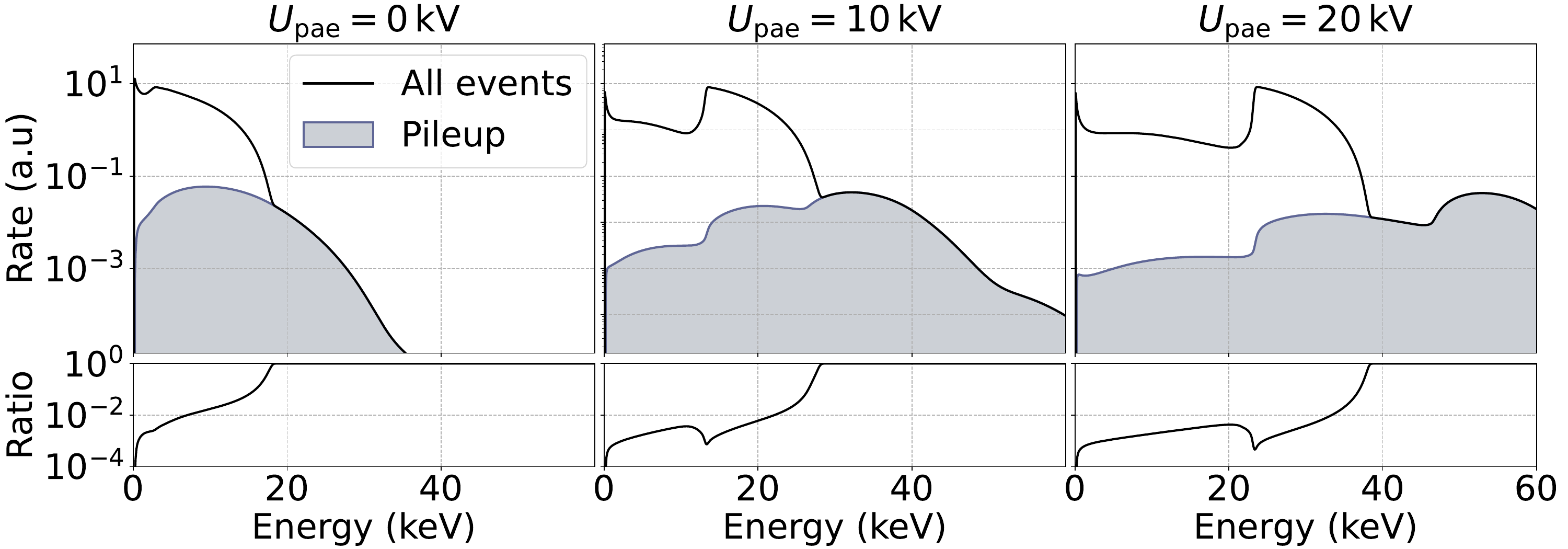}

  {\small (c) Unresolved pileup\par}

  \caption{Impact of key detector related systematic effects on the expected spectrum for different post-acceleration potentials. For pileup and backscattering, the relative contributions to the total spectrum are shown. Charge sharing does not constitute an additional spectral component but redistributes events among bins in a bin-width-dependent way; therefore, its effect is illustrated by comparing spectra before and after applying the charge-sharing response. Post-acceleration increases the electron kinetic energy and further reduces its incidence angle at the detector surface, thus lowering the backscattering probability. It also shifts unresolved pileup to energies above the endpoint and moves the signal spectrum away from regions dominated by charge-sharing effects.}
  \label{fig:systematic_effects_spectrum}
\end{figure}

Figure~\ref{fig:RW_mitigation} illustrates the strong suppression of RW backscattering achieved through the optimized magnetic-field configuration and the implementation of the new beryllium rear wall, resulting in only a small residual contribution from rear-wall events.
Relative to the nominal configuration using a gold-plated RW, the RW-induced contribution to the event rate within the fit range is reduced from $\sim$$4\times10^{-1}$ to $\sim$$4\times10^{-2}$ by magnetic-field optimization, and further to $\sim$$5\times10^{-3}$ when the beryllium RW is implemented.

\begin{table*}[t]
\centering
\caption{Systematic parameters and their associated uncertainties as considered for the sensitivity study. The last column indicates how the uncertainty assignment was obtained: data refers to constraints derived from measurements or calibration data, motivated refers to values guided by dedicated simulations and/or data–simulation comparisons, and arbitrary refers to conservative assumed values adopted where no direct constraint was available. Parameters marked unconstrained are profiled freely in the fit.}
\begin{tabular}{l|l|c|c|c|l}
\multicolumn{1}{c|}{\textbf{Category}} & 
\multicolumn{1}{c|}{\textbf{Parameter}} & 
\multicolumn{1}{c|}{\textbf{Symbol}} & 
\multicolumn{1}{c|}{\textbf{Base value}}  & 
\multicolumn{1}{c|}{\textbf{Uncertainty}} & 
\multicolumn{1}{c}{\textbf{Source}} \\ \hline \hline
\multirow{3}{*}{Source}                    & Column density total            & $\rho d$                         & $3.1 \times 10^{15}$ cm$^{-2}$\rule{0pt}{2.3ex}  & 2\%                        & data       \\ 
                                           & Column density trap             & $\rho d_{\text{trap}}$           & $3.1 \times 10^{14}$ cm$^{-2}$                   & 1\%                        & motivated  \\  
                                           & Trap magnetic field             & $B_{\text{trap}}$                & $B_{\text{src}}$ - 20\,mT                      & 3\,mT                      & motivated  \\ \hline
Rear wall                                 & Backscattering amplitude scaling factor    & $\alpha_{\text{bs}}$   & 1                                              & 10\% \rule{0pt}{2.3ex}     & motivated  \\  \hline 
\multirow{6}{*}{\shortstack[l]{Transport \\ \& Spectrometer}} & Magnetic field rear wall   & $B_{\text{rw}}$    & 0.82\,T                                        & 0.25\%\rule{0pt}{2.3ex}    & arbitrary  \\  
                                           & Magnetic field source           & $B_{\text{src}}$                 & 3.6\,T                                         & 0.25\%                     & data       \\  
                                           & Magnetic field max. downstream  & $B_{\text{md}}$                  & 3.62\,T                                        & 0.25\%                     & arbitrary       \\  
                                           & Magnetic field detector         & $B_{\text{det}}$                 & 1.41\,T                                        & 0.25\%                     & arbitrary  \\  
                                           & Retarding potential             & $U_{\text{ms}}$                  & -3.5\,kV                                       & --                         & --         \\  
                                           & Post acceleration electrode        & $U_{\text{pae}}$              & 20\,kV                                         & 1\,V                       & arbitrary  \\ \hline
\multirow{4}{*}{Detector}                  & Backscattering amplitude scaling factor   & $\beta_{\text{bs}}$    & 1                                              & 10\%                       & motivated  \\  
                                           & Dead layer parameter            & $\lambda$                        & 58\,nm                                         & 2\,nm                      & data       \\  
                                           & Charge cloud width              & $w_{\text{cc}}$                  & 20\,$\mu$m                                     & 1\,$\mu$m                  & data       \\  
                                           & Fano noise scaling factor       & $S_{\text{fano}}$                & 1                                              & 10\%                       & arbitrary  \\ \hline
\multirow{4}{*}{Readout}                   & Pulse pair resolution           & $t_{\mathrm{holdoff}}$          & 112\,ns                                        & -                          & -  \\  
                                           & Trigger deadtime                & $t_{\mathrm{pu}}$               & 1.15\,$\mu$s                                   & -                          & -  \\  
                                           & Electronic noise                & $\sigma_{\text{en}}$             & 44\,eV                                         & 10\%                       & arbitrary  \\  
                                           & Energy scale -- gain            & $G$                              & 1                                              & unconstrained              & motivated  \\  
                                           & Energy scale -- offset          & $E_{\text{offset}}$              & 0\,eV                                          & unconstrained              & motivated  \\ \hline 
Background                                 & Background rate                 & $R_{\text{bkg}}$                        & $10^{-3}$\,cps/keV                             & 10\%/keV \rule{0pt}{2.3ex} & arbitrary  \\  \hline \hline 
\end{tabular}
\label{tab:systematic_parameters}
\end{table*} 

\begin{figure}[t]
  \centering
  \includegraphics[width=\columnwidth]{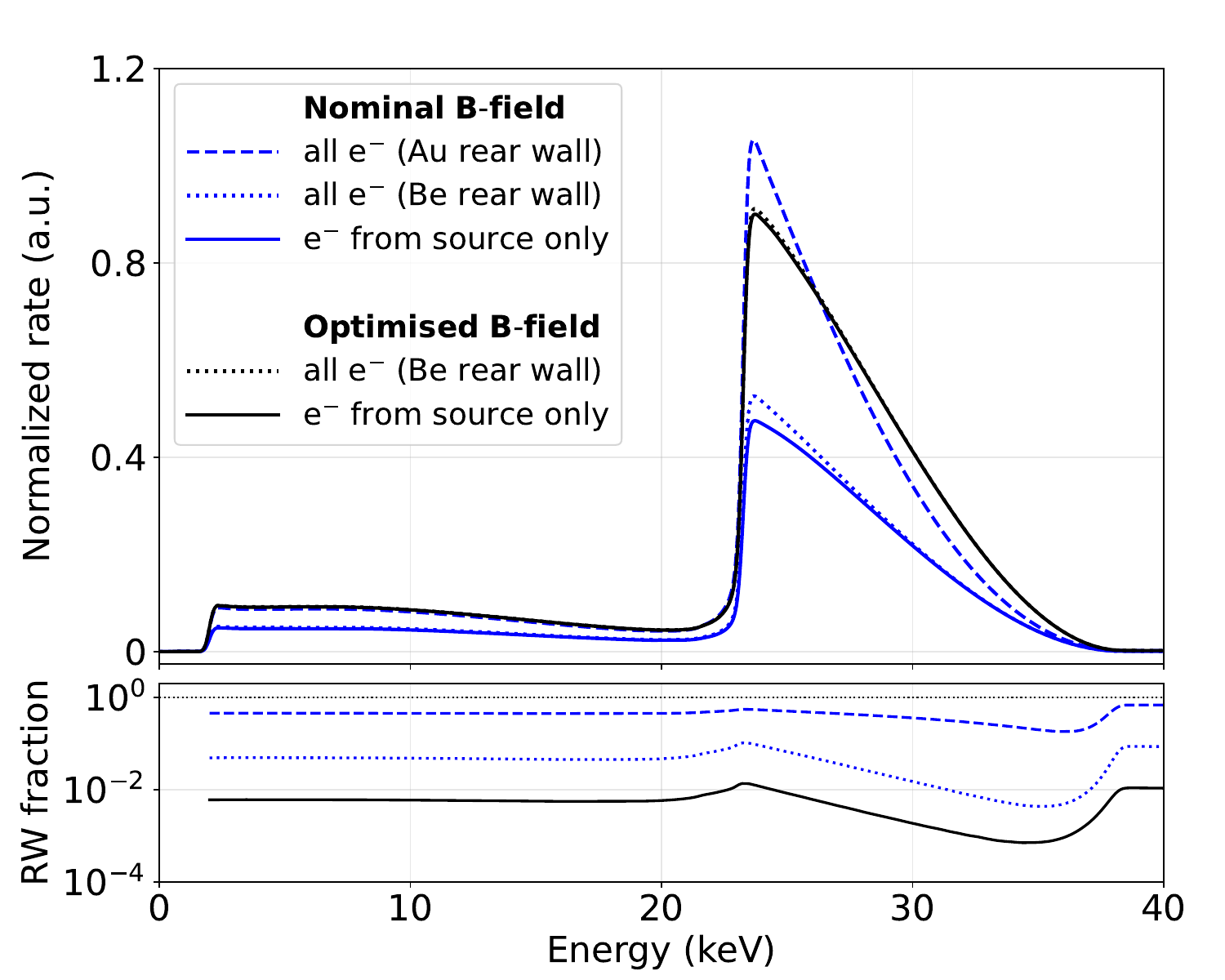}
  \caption{Mitigation of rear-wall electrons through material exchange and optimization of the electromagnetic field design. In the optimized configuration, a larger fraction of  electrons emitted in the WGTS are transported to the detector without reflections, leading to a higher source-only spectrum. 
  Replacing the gold-plated RW with beryllium significantly reduces backscattering and, together with the optimized magnetic-field configuration that reflects large-pitch-angle electrons back onto the RW, strongly suppresses rear-wall events.
}
  \label{fig:RW_mitigation}
\end{figure}

 \section{Sensitivity analysis \label{sec:5_Results}}

In this section, we present the projected sensitivity of KATRIN to keV-scale sterile neutrinos. After describing the statistical methodology, we discuss the expected statistical reach, the impact of systematic uncertainties, and the effect of potential mismodeling.

\subsection{Sensitivity Methodology}

The sensitivity results presented here were obtained using a \(\chi^2\)-based analysis of Asimov (or “reference”) datasets. The Asimov spectra are generated from the expected event rates under the null hypothesis (\(|U_{e4}|^2 = 0\)) and do not include statistical fluctuations.
At each point of a predefined two-dimensional grid covering a range of sterile neutrino masses and mixing amplitudes, the predicted spectrum for the alternative hypothesis (with parameters \(m_4\) and \(|U_{e4}|^2\)) is compared to the reference spectrum. The scalar \(\chi^2\) test statistic is then calculated as:

\begin{align}
\chi^2(|U_{e4}|^2, m_4 \mid \vec{p}) \;=\;&\;
    \bigl[\vec{\Gamma}(|U_{e4}|^2, m_4 \mid \vec{p}) - \vec{\Gamma}_{\text{ref}}\bigr]^{T} \nonumber \\
& V^{-1} \;
    \bigl[\vec{\Gamma}(|U_{e4}|^2, m_4 \mid \vec{p}) - \vec{\Gamma}_{\text{ref}}\bigr],
\label{eq:chi2_expanded}
\end{align}
where \(\vec{\Gamma}_{\text{ref}}\) is the Asimov spectrum without sterile mixing and \(\vec{\Gamma}(|U_{e4}|^2, m_4 \mid \vec{p})\) is the spectrum including a sterile component. \(\vec{p}\) represents any relevant nuisance parameters, and \(V\) is the covariance matrix. 
\(V\) can be set to include purely statistical uncertainties (\(V^{\text{stat}}\)) or the combination of statistical and systematic uncertainties \(\bigl(V^{\text{stat}} + V^{\text{syst}}\bigr)\), depending on whether systematic uncertainties are considered. 
The statistical covariance is calculated assuming Poisson counting in each bin and given by:
\[
V^{\text{stat}}_{ii} = \Gamma_{\text{ref},i}, 
\qquad
V^{\text{stat}}_{ij} = 0 \quad (i \neq j).
\]
The impact of systematic uncertainties is assessed by generating spectra in which the input parameters are randomly varied according to Gaussian distributions with standard deviations given by their respective uncertainties. For each random draw of the systematic parameters, a new spectrum \(S^k\) is generated. By accumulating \(N\) such simulated spectra, a bin-to-bin covariance matrix is computed as
\begin{equation}
    V_{ij}^{\text{syst}} \;=\;
    \frac{1}{N} \sum_{k=1}^{N}
    \bigl(S_i^k - \overline{S}_i \bigr)\,
    \bigl(S_j^k - \overline{S}_j \bigr),
\end{equation}
where \(S_i^k\) is the content of the \(i\)-th energy bin for the \(k\)-th sample, and \(\overline{S}_i\) is the mean bin content over all samples. 
Sensitivities are calculated using fits in which two key parameters, the global normalization (source activity) and the pileup amplitude, are allowed to vary freely. This ensures that any global rate effect is absorbed. Most other systematic effects are incorporated via precomputed covariance matrices, 
while a small subset of nuisance parameters are profiled explicitly when indicated.
By contrast, explicit profiling over a large set of nuisance parameters would be more computationally demanding. 
Using a covariance-based approach greatly reduces calculation time and was observed to yield slightly more conservative limits than full profiling.

The sterile-neutrino parameter space is explored by scanning the ranges \(m_4 \in [0, 15]\,\text{keV}\) linearly and \(|U_{e4}|^2 \in [10^{-7}, 0.5]\) logarithmically. Since the tritium $\upbeta$-decay spectrum is shifted by \(20\,\text{keV}\) due to the post-acceleration potential, and a retarding potential of \(U_{\text{ms}} = -3.5\,\text{kV}\) is applied, the energy of the measured electrons of interest span from \(23.5\,\text{keV}\) up to the tritium endpoint at approximately \(38.6\,\text{keV}\). The \(\chi^2\) fit is performed over this  
range, using a bin width of \(100\,\text{eV}\). At each grid point, the \(\chi^2\) value is evaluated, and sensitivity contours are derived under the assumption of Wilks’s theorem and are reported at the 95\% CL (\(\chi^2_{\text{crit}} = 5.99\)).

\subsection{Statistical sensitivity \label{subsec:statistic}}

The expected statistical sensitivity of KATRIN to keV sterile neutrinos for various measurement times is presented in Fig.~\ref{fig:TRISTAN_sensitivity_stat_time}. The sensitivity curves exhibit a characteristic shape over the accessible mass range. It is driven by the tritium $\upbeta$-spectrum and the mass-dependent position of the sterile-neutrino signature, with the optimal sensitivity reached at the midpoint of the mass range.
\begin{figure}[t]
  \centering
  \includegraphics[width=\columnwidth]{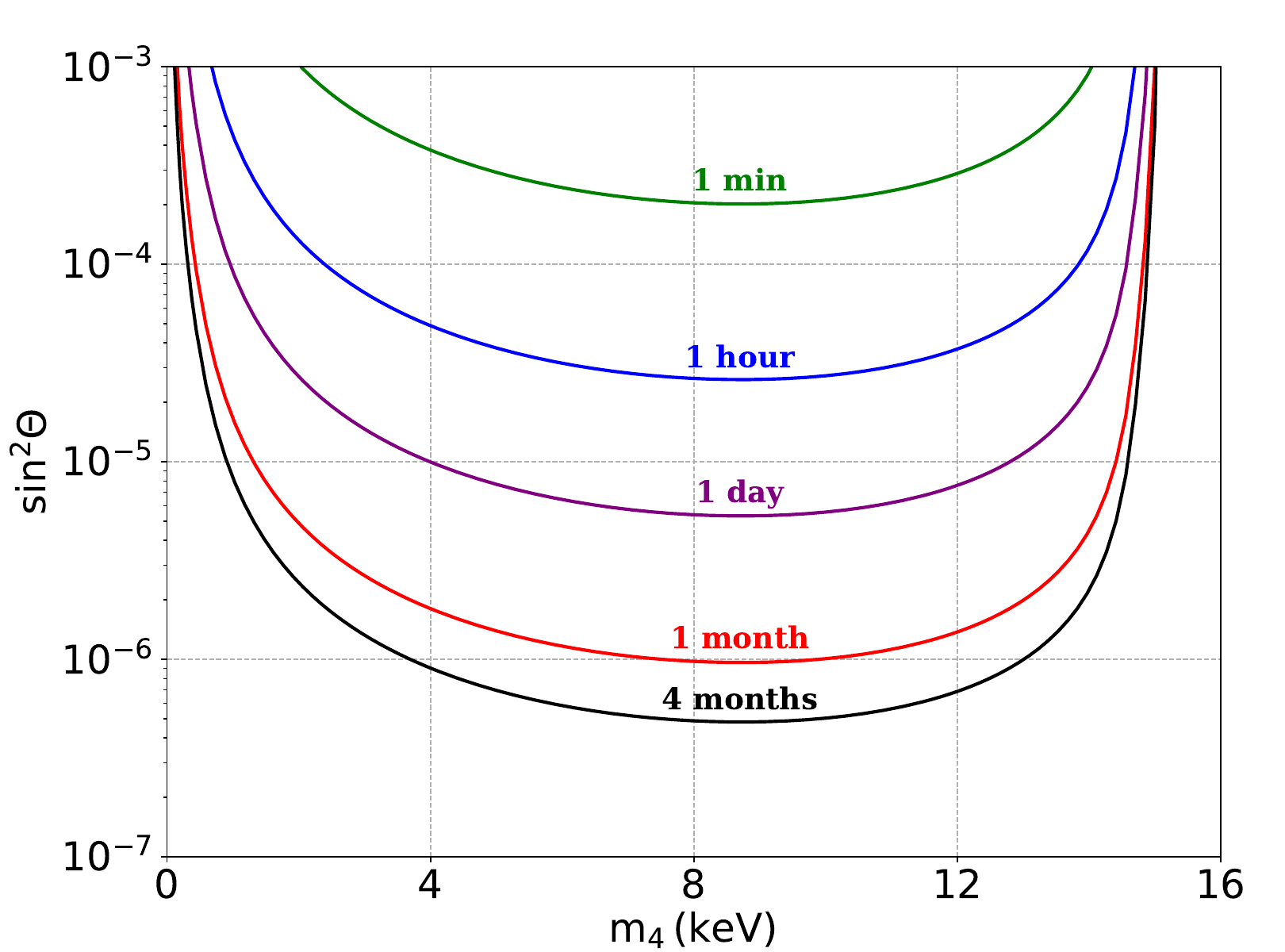}
  \caption{Impact of the measurement time on statistical sensitivity. For all cases, a common retarding potential of \(U_{\text{ms}} = -3.5\,\text{kV}\) as well as a common column density of \( \rho d = 0.6\% \) are used.}
  \label{fig:TRISTAN_sensitivity_stat_time}
  \vspace{1em} 
  \includegraphics[width=\columnwidth]{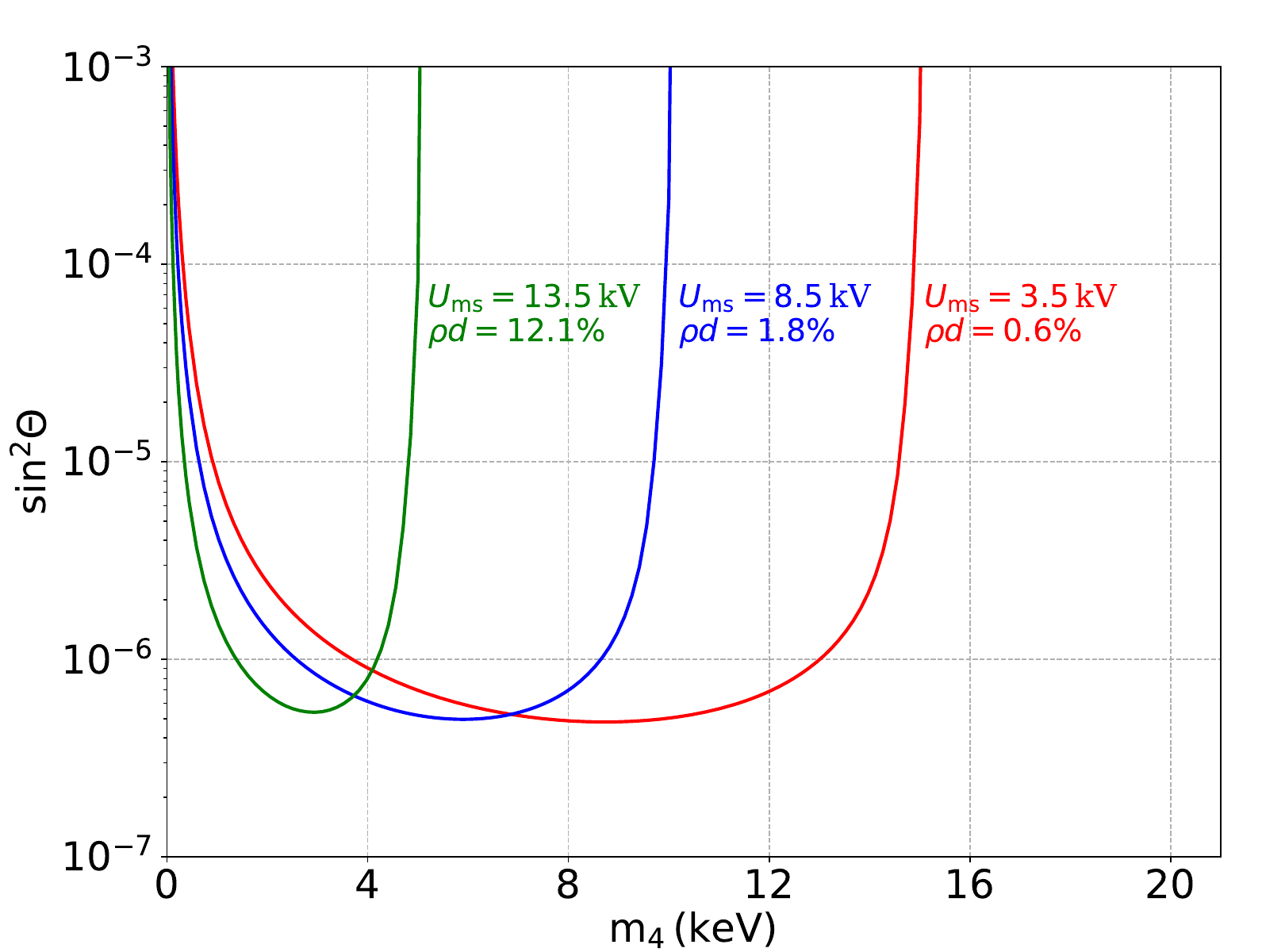}
  \caption{Impact of the retarding potential on statistical sensitivity. For all cases, a measurement time of 4 months was considered. The column density was tuned to yield a rate of 100\,kcps/pixel.}
  \label{fig:TRISTAN_sensitivity_stat_qU}
\end{figure}
Thanks to a high input count rate of $\sim$100\,kcps per pixel, KATRIN can collect $N_{\text{evt}}\sim1.5\times10^{11}$ events and reach a statistical sensitivity of 
$|U_{e4}|^2 \sim 2\times10^{-5}$ at the midpoint of its mass range after just one hour of data collection. 
Because statistical uncertainties scale as $\sqrt{N_{\mathrm{evt}}}$ according to Poisson statistics, the statistical sensitivity improves as $\propto 1/\sqrt{T_{\text{meas}}}$, where $T_{\text{meas}}$ is the total measurement time. Consequently, KATRIN has the potential to reach a statistical limit of up to $\sim$$5\times10^{-7}$ with four months of data taking, corresponding to a total accumulated statistic of about $4\times10^{14}$ events.

The tritium $\upbeta$-decay spectrum theoretically enables the search for sterile neutrinos with masses extending up to the tritium endpoint energy of 18.6\,keV. In practice, however, systematic effects with strong contributions at low energy, such as backscattering at the RW and detector or potential non-adiabatic transport through the main spectrometer, may necessitate operating at a higher retarding potential. This would truncate the low-energy part of the spectrum and thereby reduce the accessible sterile-neutrino mass range. The impact of the retarding potential on the accessible mass range is illustrated in Fig.~\ref{fig:TRISTAN_sensitivity_stat_qU} for several $qU_{\text{ms}}$ values: 1.5, 3.5, 8.5 and 13.5\,keV respectively corresponding to accessible mass ranges of 17, 15, 10 and 5\,keV.
Although increasing the retarding potential may be necessary to mitigate challenging low-energy systematic effects, high statistical sensitivity can be maintained by adjusting the column density to keep an input detector rate of 100 kcps per pixel.

\subsection{Systematic uncertainties \label{subsec:systematics}}
\subsubsection{Assigned systematic uncertainties}

A comprehensive set of systematic effects has been evaluated to assess their impact on the experimental sensitivity. The corresponding nuisance parameters, their reference values, and assigned uncertainties are summarized in Table~\ref{tab:systematic_parameters}. The adopted uncertainties are based on a combination of beamline measurements carried out for the neutrino-mass measurement, dedicated measurements with the TRISTAN detector and DAQ system, or based on conservative assumptions when data were not available.

Source-related uncertainties include the tritium column density in the WGTS and magnetic trapping effects. A relative uncertainty of 2\% is assigned to the column density, larger than the 1\% used in the neutrino-mass analysis~\cite{KATRIN_NuMass3_Science_2025} due to the reduced statistical precision expected from electron-gun calibration measurements at low column densities.
Magnetic trapping is parameterized by an effective trap column density and an average trap magnetic field, with uncertainties of 1\% and 0.15\%, respectively, motivated by dedicated simulation studies.

An energy- and angle-independent relative uncertainty of 10\% is assigned to the electron backscattering probability at the rear wall to account for residual modeling uncertainties. This choice is motivated by comparisons between measured and simulated backscattering coefficients with the TRISTAN detectors~\cite{TRISTAN_Backscattering_Spreng_JINST_2024, TRISTAN_Backscattering_Nava_JPCS_2021}, as well as on published comparisons of experimental data with \textsc{Geant4} simulations for relevant materials and energies~\cite{Sim_Geant4BackscatterValidation_Sung_IEEE_2015,
Sim_Geant4Backscattering_Dondero_NIMB_2018}.

A relative uncertainty of 0.25\% was previously measured for the source magnetic-field~\cite{KATRIN_PhDThesis_Trost_2019}. 
As a baseline, the same 0.25\% relative uncertainty is considered for the source, rear-wall, maximal downstream and detector magnetic fields. Uncertainties associated with the main spectrometer retarding potential and the post-acceleration energy have been demonstrated to be stable and reproducible at the sub-ppm level~\cite{KATRIN_RetardingVoltage_Rodenbeck_JINST_2022}. For this study, the post-acceleration energy uncertainty was set to an arbitrary value of 1\,eV.

Regarding detector related effects, a 3.5\% uncertainty is assigned to the dead-layer width parameter and a 5\% uncertainty to the charge-cloud width, based on  data taken with the TRISTAN detectors as discussed in Sec.~\ref{sec:4C_Input}. 
Relative uncertainties of 10\% are assigned to both the Fano noise and the electronic noise. This choice conservatively covers the resolution spread observed in detector characterization measurements at different energies.
Unmodeled backreflection of electrons backscattered at the detector surface is covered by assigning a  10\% uncertainty to the detector backscattering probability, motivated by dedicated simulations.
DAQ calibration parameters, including the gain and offset, are not constrained through covariance matrices but are instead profiled as free parameters in the fit to account for residual energy-scale uncertainties. This conservative treatment allows the fit to self-calibrate the global energy scale using the high event statistics, thereby absorbing potential small residual mis-calibrations without biasing the sensitivity. Dedicated calibration measurements with the TRISTAN detector and the DAQ system are ongoing to independently constrain these parameters and reduce their impact in future analyses. 
DAQ non-linearities, primarily originating from the analog-to-digital conversion of detector signals, are accounted for in the sensitivity analysis. Detailed measurements and simulations have shown that these effects induce smooth and energy-dependent distortions of the reconstructed spectrum. When mitigation techniques are applied, their impact on the sterile-neutrino sensitivity is negligible~\cite{TRISTAN_DAQ_Gavin_arXiv_2026}.

Finally, to account for possible deviations from the simplified flat background model, a conservative uncorrelated bin-to-bin uncertainty of 10\% of the background rate per 1\,keV is assigned.

\subsubsection{Systematic impact on the sensitivity}

The relative impact of individual systematic uncertainties on the measured spectrum within the analyzed energy range is illustrated in Fig.~\ref{fig:Systematic_uncertainty}. 
\begin{figure}[tbp]
  \centering
 \includegraphics[width=\linewidth]{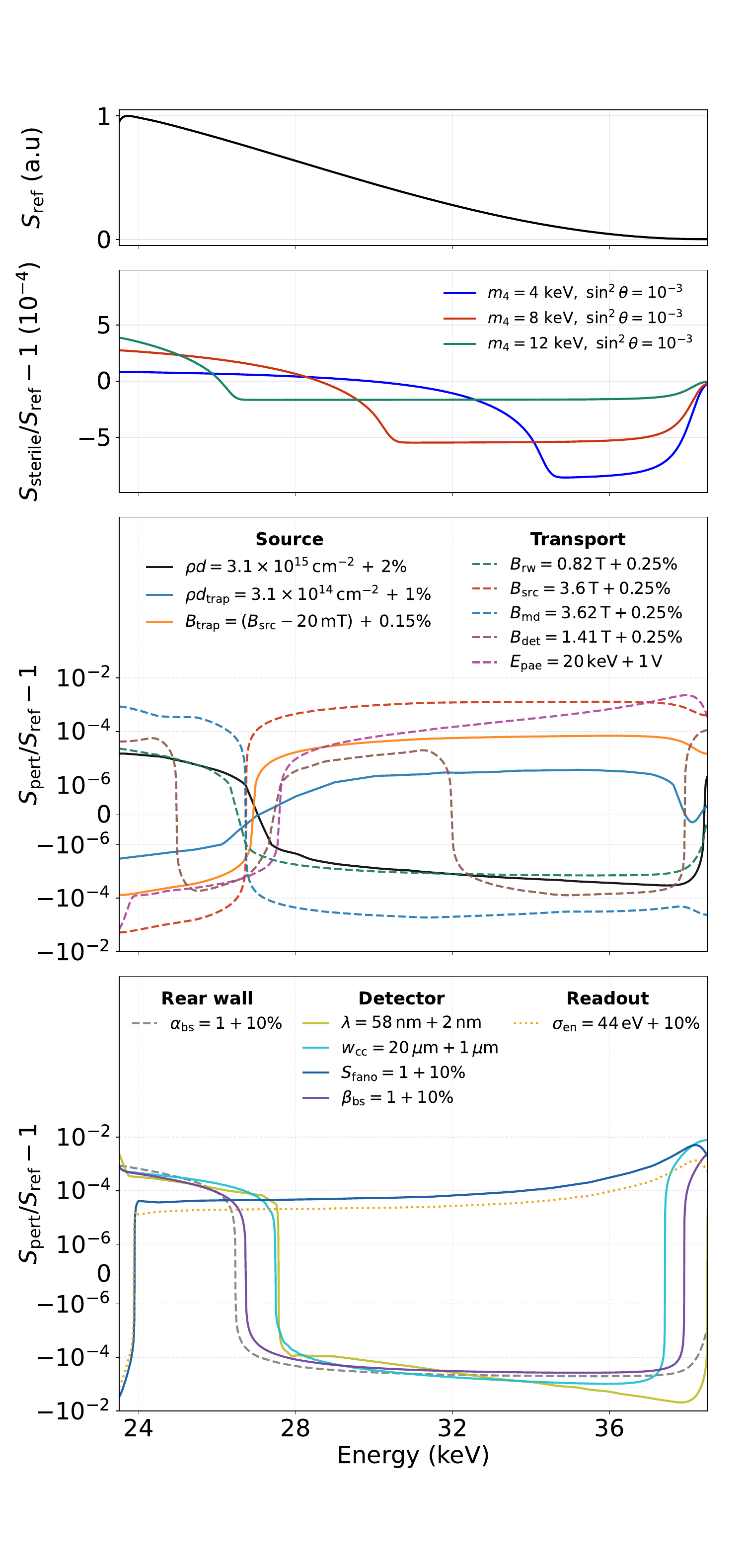}
\caption{
Impact of a sterile-neutrino signal and individual systematic uncertainties on the measured spectrum
within the region of interest.
(top) Reference expected spectrum $S_{\mathrm{ref}}$.
(upper middle) Relative spectral distortion induced by a sterile-neutrino admixture for three representative
sterile-neutrino masses $m_4 = 4,\;8,\;12~\mathrm{keV}$ at fixed mixing $\sin^2\theta = 10^{-3}$.
(lower middle and bottom) Relative spectral distortions induced by individual systematic variations applied one at a time
around the reference model. The considered effects are grouped by category and correspond to the parameter variations listed in Table~\ref{tab:systematic_parameters}. The lower-middle panel shows source and transport effects, while the bottom panel shows rear-wall, detector, and readout effects. In all ratio panels, only shape distortions are shown, with the overall normalization fixed to that of the reference spectrum.
}
  \label{fig:Systematic_uncertainty}
\end{figure}
Most systematic effects introduce smooth, energy-correlated distortions with amplitudes typically well below the percent level. Their spectral behavior is qualitatively distinct from the characteristic distortion induced by a keV-scale sterile neutrino, which features a kink-like spectral feature at $E = E_0 - m_4$ accompanied by an extended distortion at lower energies.
The impact of systematic uncertainties on the sterile-neutrino sensitivity is summarized in Fig.~\ref{fig:TRISTAN_sensitivity_placeholder} for each group of systematic effects, while the individual contributions are shown in Fig.~\ref{fig:Sens_collab_20.03.2025_2x2}. Starting from the statistics-only sensitivity, systematic effects are added sequentially at the group level and finally combined all together.
When all systematic uncertainties are combined, the projected sensitivity is degraded by a factor ranging from 10 to 50 with respect to the statistical limit. 
For a reference data-taking period of four months, this corresponds to an expected sensitivity in the order of  $|U_{e4}|^2 \sim 2\times10^{-5}$ over the sterile-neutrino mass range $m_4 \simeq (4$--$13)~\mathrm{keV}$.
Repeating the analysis for different measurement times shows that the relative loss compared to the statistics-only sensitivity remains consistent.
Importantly, the achievable sensitivity is not dominated by any single group of systematic effects, but rather by the cumulative impact of the several sub-dominant contributions.
Source\mbox{-,} transport\mbox{-,} and detector-related uncertainties lead to comparable sensitivity loss, with transport effects exhibiting a slightly larger impact on average. 
When considered individually, each of these groups still allows sensitivities typically better than $5\times10^{-6}$. 
Rear-wall and DAQ-related systematics are found to have a noticeably smaller impact on the sensitivity.
At the level of individual nuisance parameters, a direct correspondence between the spectral distortion induced by a given systematic effect and its impact on the sterile-neutrino sensitivity is typically not straightforward to identify. However, a clear trend emerges: systematic effects that predominantly modify the low-energy part of the reconstructed spectrum (e.g. RW backscattering, charge-sharing cloud width) tend to induce a larger degradation of the sensitivity at higher sterile-neutrino masses. 

Within the source-related systematics, the sensitivity is dominated by uncertainties associated with the simplified modeling of magnetic trapping in the WGTS, in particular the effective magnetic-field strength inside the trap regions. 
This highlights the importance of further experimental and simulation studies of this effect. 
Sub-dominant but still non-negligible contributions are obtained from uncertainties of the total column density and the fraction of magnetically trapped tritium.

For beamline and transport effects, the sensitivity loss is primarily driven by the uncertainty in the source magnetic field, which controls electron transport both upstream and downstream of the source. 
Uncertainties in the maximal downstream field and the rear-wall magnetic fields produce effects of comparable but smaller magnitude. 
By contrast, the detector magnetic field and the PAE energy have a non-negligible but much smaller impact on the sensitivity.

Detector-related systematic uncertainties show no single dominant contribution. 
Dead-layer width, charge-sharing cloud width, and detector backscattering all produce sensitivity degradations of similar magnitude, while the impact of the intrinsic energy-resolution uncertainty is found to be negligible.

The sensitivity loss due to DAQ-related effects is dominated by the energy-scale calibration parameters.
Uncertainties associated with electronic noise, pulse-pair resolution, and DAQ non-linearity have at most a marginal impact on the sensitivity.

Finally, uncertainties associated with the background spectral shape are found to have no measurable impact on the sensitivity and are therefore not shown in the figures.
 
\begin{figure}[t]
  \centering
{
   \includegraphics[width=\linewidth]{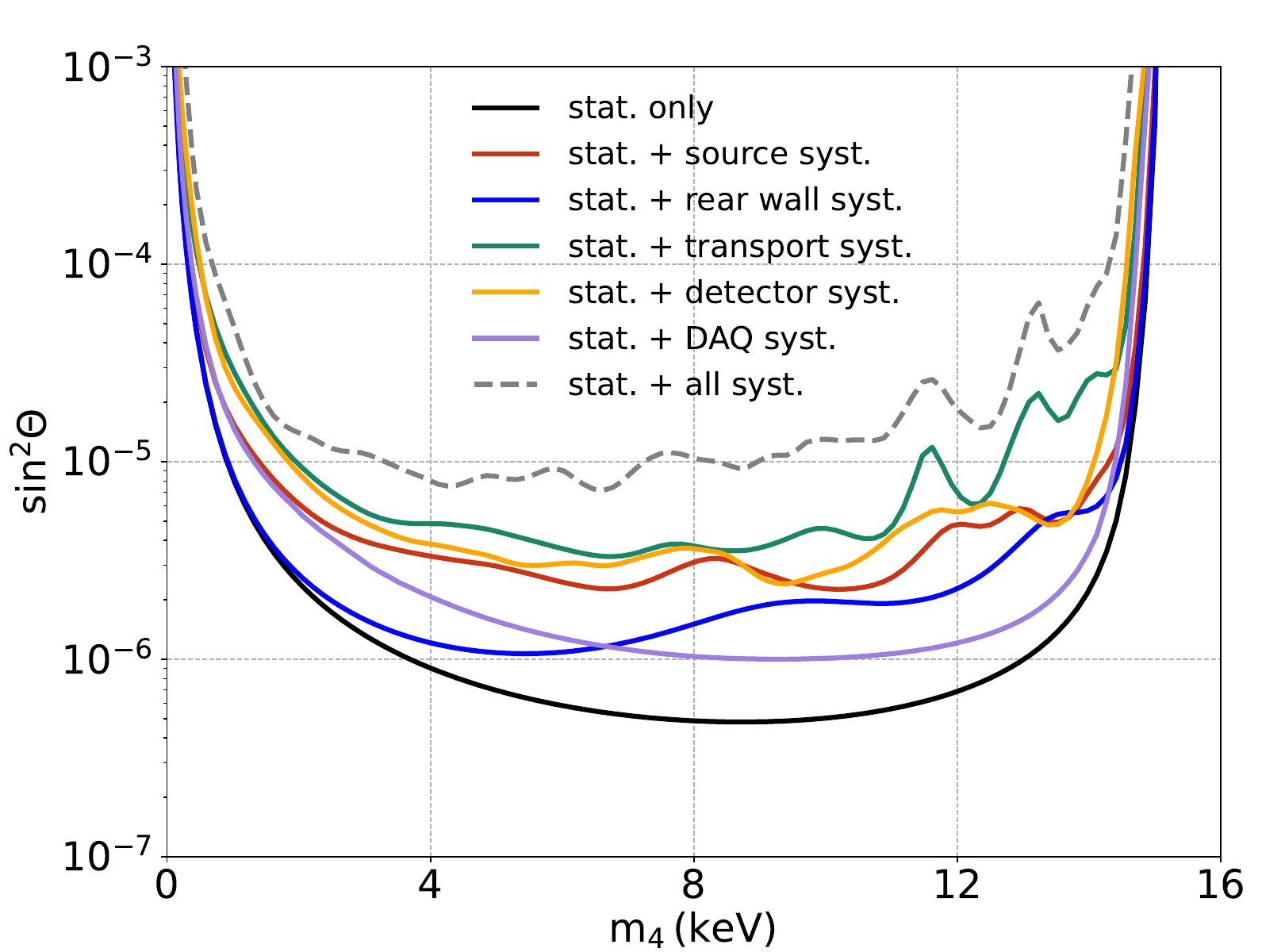}
  }
  \caption{Impact of systematic uncertainties on the projected KATRIN sensitivity to keV-scale sterile neutrinos for four month of detector live-time. The black solid curve shows the statistics-only sensitivity. Colored solid curves represent the sensitivity obtained when adding, on top of statistical uncertainties, one category of systematic effects at a time as reported in Table~\ref{tab:systematic_parameters}.
  The gray dashed curve shows the sensitivity combined for all systematic uncertainties.
}
  \label{fig:TRISTAN_sensitivity_placeholder}
\end{figure}
\begin{figure}[t]
  \centering
    \includegraphics[width=\linewidth]{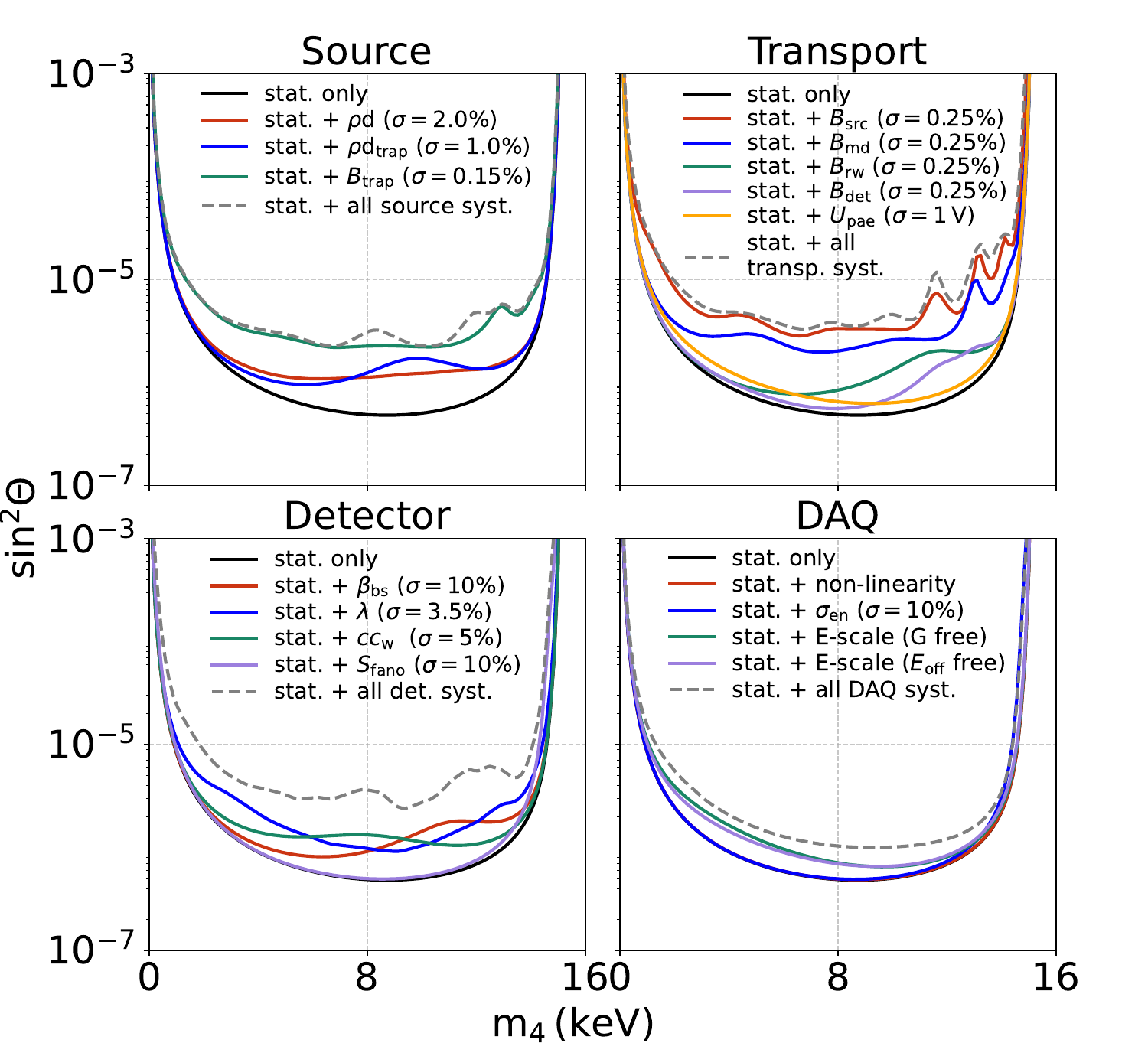}
  \caption{Breakdown of the impact of individual systematic uncertainties on the sterile-neutrino sensitivity.
Each panel shows the statistics-only sensitivity (black solid line) and the sensitivity obtained when adding a single systematic uncertainty at a time, grouped by category.
The gray dashed curves show the combined impact of all systematic uncertainties within each category.
}
  \label{fig:Sens_collab_20.03.2025_2x2}
\end{figure}

\subsection{Modeling inaccuracy}

In the previous section, we propagated uncertainties of known systematic effects by varying model parameters within their assumed range.  These systematics generally introduce smooth and energy-correlated distortions to the spectrum that do not mimic the distinctive signature of a keV sterile-neutrino signal. Because of this, marginalizing over them retains most of the sensitivity to the signal.
However, this approach implicitly assumes that the model parametrization of each systematic effect is accurate. While we tested parameter uncertainties within a correct model, we did not evaluate the impact of a potential mismodeling. If the true spectrum deviates from the assumed parametrization, such deviation could degrade the sensitivity in ways not addressed by the current study.

In this section, we investigate the impact of potential mismodeling by introducing a shape uncertainty on a known systematic contribution. Similar investigations of spectral mismodeling have been performed previously~\cite{TRISTAN_Concept_Mertens_JCAP_2015}, where generic shape distortions were introduced to assess their impact on sterile-neutrino searches. 
Here, we instead focus on the specific and physically motivated systematic effect associated with RW backscattering.
Concretely, an uncorrelated bin-to-bin uncertainty is applied to the RW spectral component, and the magnitude of this uncertainty is varied to assess its impact on the projected sensitivity.
The results of this study are shown in Fig.~\ref{fig:TRISTAN_shape_sensitivity}. The impact of a normalization uncertainty on the RW backscattering contribution is compared to that of a shape uncertainty, whose amplitude is varied between benchmark values of $0.01\%$ and $10\%$ per keV.
\begin{figure}[t]
  \centering
   {
    \includegraphics[width=\linewidth]{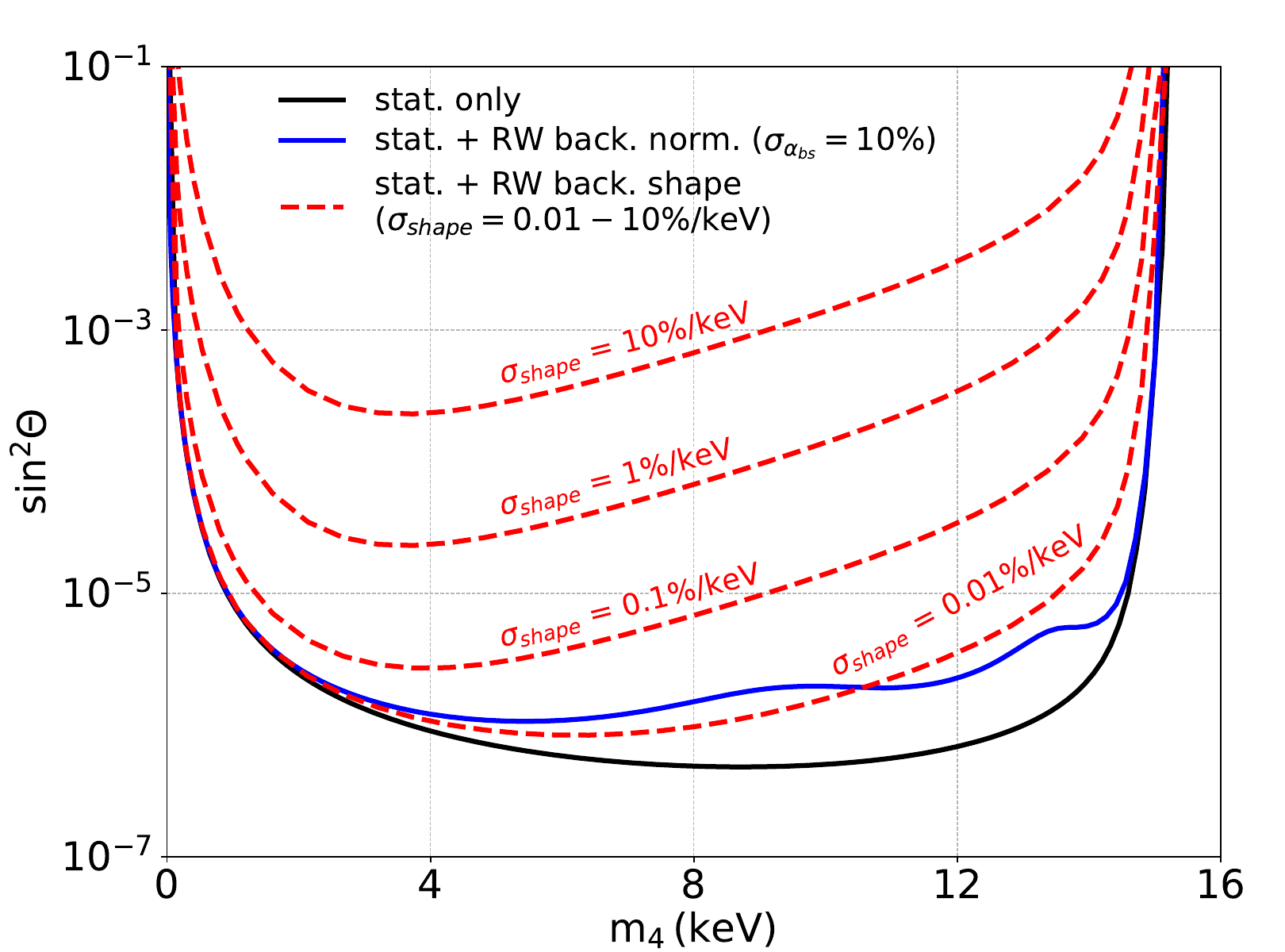}
  }
  \caption{
  Impact of normalization and shape uncertainties of the Be rear-wall backscattering contribution on the sterile-neutrino sensitivity. The blue solid line corresponds to a normalization uncertainty, while the dashed lines show the effect of shape uncertainties modeled as uncorrelated bin-to-bin uncertainties in the range of $0.1\%$--$10\%$/keV, used as benchmark values.
  }
  \label{fig:TRISTAN_shape_sensitivity}
\end{figure} 
Several conclusions can be drawn from this study. First, mismodeling of the RW backscattering spectral shape can substantially degrade the sterile-neutrino sensitivity, with an impact that far exceeds that of a normalization uncertainty alone. When only a normalization uncertainty is considered, the sensitivity remains largely preserved, with losses below a factor of four over the full sterile-neutrino mass range relative to the statistics-only case.
In contrast, a shape uncertainty at the level of $\lesssim 0.01\%$ per keV is required to induce a comparable loss in sensitivity. Increasing the shape uncertainty by one order of magnitude leads to a further order-of-magnitude reduction in sensitivity. For a benchmark shape uncertainty of $1\%$ per keV, the sensitivity is degraded by factors ranging from $\mathcal{O}(10)$ at low sterile-neutrino masses up to $\mathcal{O}(10^{3})$ at higher masses. 
This mass dependence results from the increasing relative contribution of RW backscattering events at lower electron energies (see Fig. \ref{fig:RW_mitigation}). Because this study assumes identical uncorrelated relative uncertainties per bin across the entire energy range, mismodeling effects become more pronounced in the low-energy region, leading to a stronger degradation of the sensitivity at higher sterile-neutrino masses.
At an intermediate mass of $m_4 \simeq 7.5~\mathrm{keV}$, the sensitivity deteriorates from $\sin^2\theta \sim 5\times10^{-7}$ to $\sim$$5\times10^{-5}$.
  
It is important to emphasize that this study does not aim to model a specific physical mismodeling mechanism, but rather provides a conservative stress test of the sensitivity to generic shape uncertainty affecting a dominant systematic component. In practice,  fully uncorrelated bin-to-bin shape uncertainties of this magnitude are not expected from the underlying physics of electron transport or scattering. Similar considerations apply to other systematic effects, which are generally driven by smooth and correlated energy-dependent distortions. 
Overall, these results highlight the critical importance of accurate modeling of systematic effects. 
They also underscore the need for in situ measurements with the TRISTAN detector to experimentally validate and characterize these contributions in order to achieve optimal sensitivity.

\subsection{Comparison with existing constraints}

The projected KATRIN sensitivity can be placed in a broader experimental context by comparing it with existing laboratory limits and astrophysical constraints in the $(m_4, |U_{e4}|^2)$ parameter space, as shown in Fig.~\ref{fig:keV_limits}.
\begin{figure}[t]
\centering
{
 \includegraphics[width=\linewidth]{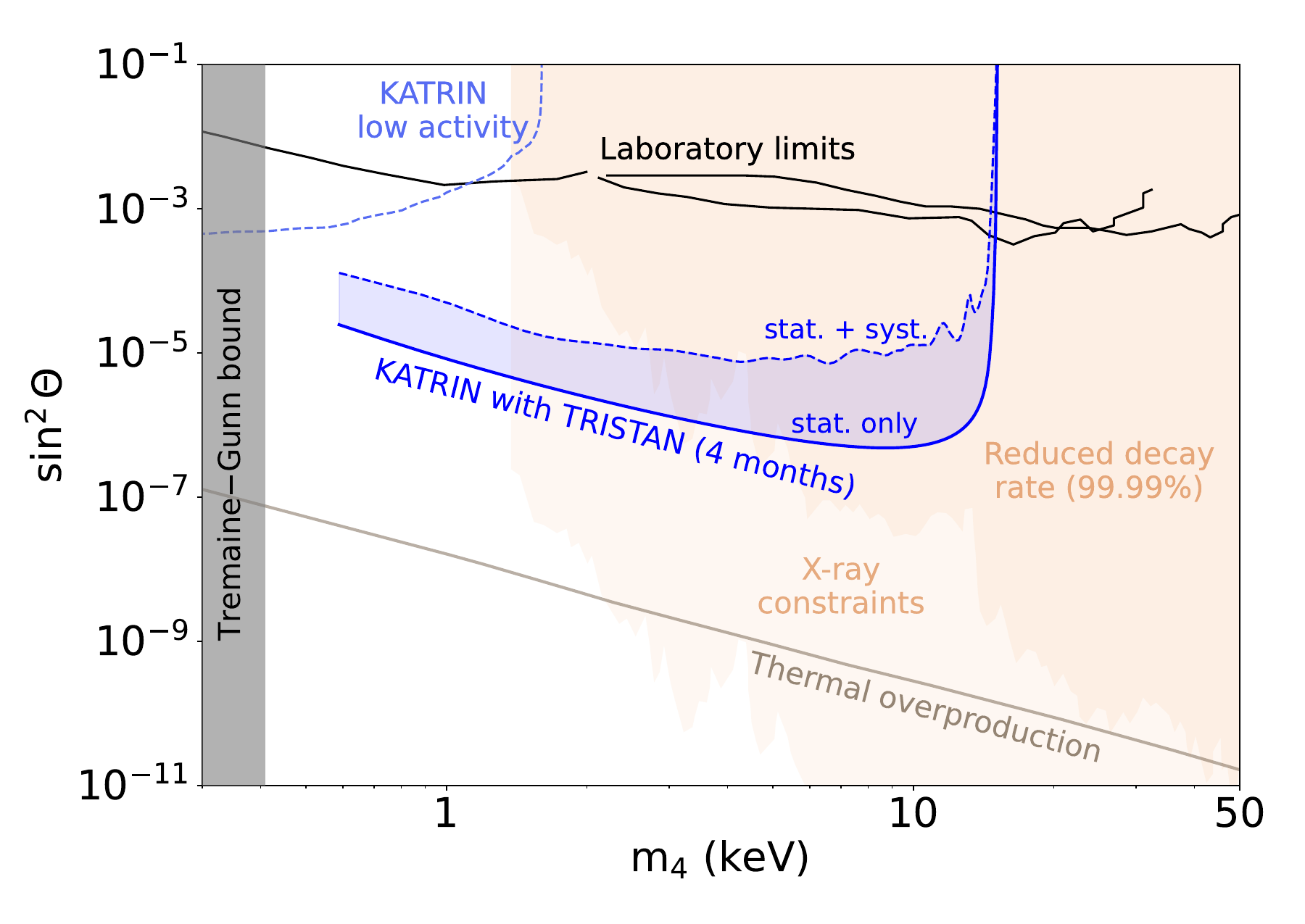}
}
\caption{Projected KATRIN sensitivity to keV sterile neutrinos with the TRISTAN detector upgrade.
The expected exclusion sensitivity obtained after 4 months of data taking is shown for a statistics-only analysis (solid blue line) and including systematic uncertainties (dashed blue line), as summarized in Table~\ref{tab:systematic_parameters}. 
Existing laboratory constraints in the mass range (0.5--50)\,keV are shown as a dashed light-blue line for the KATRIN low-activity measurement~\cite{KATRIN_keVSterile_EPJC_2023} and as solid black lines for other experiments~\cite{SterileSearch_Troitsk2_JETPL_2017,SterileSearch_35S_Holzschuh_PLB_2000,SterileSearch_63Ni_Holzschuh_PLB_1999}.
Selected astrophysical constraints, including X-ray decay limits (light-orange shaded region) and bounds from thermal overproduction (solid medium brown) taken from Ref.~\cite{KATRIN_DMProspects_PRD_2019}, are shown for comparison. A model-dependent relaxation of the X-ray decay limits is indicated by the darker light-orange shading. The grey band at low masses denotes the phase-space region excluded by the Tremaine--Gunn bound~\cite{SterileDM_PhaseSpace_TremaineGunn_PRL_1979}.
}
  \label{fig:keV_limits}
\end{figure}
The projected KATRIN sensitivity with the TRISTAN detector extends well beyond previous searches performed by KATRIN and other existing laboratory bounds. While current laboratory constraints typically reach the $\mathcal{O}(10^{-3})$ level in mixing, KATRIN has the potential to probe mixing amplitudes down to $\mathcal{O}(10^{-5})$ over the sterile-neutrino mass range $m_4 \simeq (4$--$13)~\mathrm{keV}$. KATRIN will thus provide a direct laboratory test that is independent of cosmological and astrophysical assumptions, and complementary to indirect astrophysical searches.

Other laboratory experiments probing the same parameter space utilize magnetic microcalorimeters (MMCs) to study the $\upbeta$-decay spectra of $^3\mathrm{H}$ and $^{241}\mathrm{Pu}$~\cite{SterileSearch_LiFESNS_arXiv_2026,SterileSearch_MAGNETONU_PRC_2026}. 
MMC-based measurements benefit from the energy resolution of cryogenic detectors and from the radioactive source being embedded in the detector itself, thereby avoiding systematic effects associated with beamline-based measurements. 
Nonetheless, individual MMCs are typically limited to activities of $\mathcal{O}(100~\mathrm{Bq})$. As a result, year-scale measurements with such technologies are projected to reach statistical sensitivities to mixing in the range $10^{-2}$--$10^{-4}$ for sterile-neutrino masses from 1 to 18\,keV.
The projected KATRIN sensitivity, including systematic effects, also remains competitive in the context of proposed searches based on $^3\mathrm{H}$ nanospheres, which still require further technological development and longer measurement times~\cite{SterileSearch_QuantumSensors_Carney_PRXQuantum_2023}.

\subsection{Discussion}

In this work, we have evaluated the sensitivity of the KATRIN experiment with the TRISTAN detector to keV-scale sterile neutrinos using a forward-convolution framework that incorporates the dominant experimental systematic effects. The results demonstrate strong sensitivity well beyond current laboratory limits. To put these results into perspective, we now discuss two key limitations of this study.

\textit{Model limitations --}
The convolution-based framework relies on precomputed response matrices defined on discretized energy and pitch-angle grids. Its accuracy is therefore governed by an intrinsic compromise between binning granularity, memory usage, and computational time, such that high numerical accuracy is rapidly limited by technical constraints. 
In addition, the code does not account for the potential backreflection of electrons backscattered at the detector. It also does not include an explicit radial dependence of the effects, and therefore cannot fully capture spatially resolved effects. More generally, the modular treatment of individual systematic effects makes it challenging to accurately model complex coupled processes, particularly when multiple interactions occur sequentially or when correlations depend on timing or spatial information.
Furthermore, several effects are treated using simplified parameterizations, as discussed previously. This includes, for example, the modeling of magnetic trapping in the source approximated as regions of constant magnetic field, charge sharing without explicit treatment of triple-pixel events, and unresolved pileup assumed to be an energy independent process. For such effects, the set of parameters varied in this study may not fully encompass all possible spectral distortions and the resulting sensitivity loss.

To complement the approach presented here, a full event-by-event Monte Carlo simulation of the KATRIN beamline equipped with the TRISTAN detector is under active development. It combines KASSIOPEIA's accurate electromagnetic-field modeling with the robust \textsc{Geant4} framework for simulating electron interactions with matter. This code tracks individual electrons from their points of production in the tritium source to their final energy depositions in the detector. At each stage of propagation, the electron energy, pitch angle, time-of-flight, and spatial coordinates are recorded, enabling a detailed reconstruction of successive interactions and transport histories. 
By explicitly tracking timing and spatial information, this framework naturally accounts for coupled effects such as detector backscattering followed by backreflection and fully captures spatial effects, including pixel changes between successive energy depositions caused by backreflection and electron drift.
At present, the computational demands of this full Monte Carlo approach, together with its ongoing development status, prevent its use for large-scale sensitivity scans of the type presented here. Nevertheless, it is expected to become an essential component of the TRISTAN analysis framework, providing refined modeling of the systematic effects and support for the calibration analyses.

\textit{Challenge of robust modeling --}
We have shown that the sensitivity to keV-scale sterile neutrinos relies critically on precise modeling of systematic effects. While the present simulation framework provides a robust tool for sensitivity evaluation and systematic exploration, achieving the projected sensitivity in experimental data will be challenging. As demonstrated in this work, mismodeling of dominant systematic contributions has the potential to significantly degrade the sensitivity. It is therefore critical for this next phase of KATRIN to (i) optimize the beamline configuration to minimize the impact of systematic effects, (ii) develop accurate models that account for potentially correlated and time-dependent effects approximated or neglected in the present work, and (iii) perform extensive experimental validation and calibration of the dominant response components.
Owing to its unique experimental configuration, KATRIN will have the opportunity to measure ultra-high-statistics spectra and to operate the beamline under varied electromagnetic settings. This flexibility enables systematic effects to be studied both through mitigation and deliberate enhancement, providing unique cross-checks of the modeling. 
In addition, TRISTAN will leverage ongoing calibration measurements performed with a dedicated experimental setup at KIT, where detector modules are characterized in a replica of the beamline detector section operated under conditions similar to those of the experiment, thereby providing essential input for validating and refining the model prior to detector installation.

 \section{Conclusion \label{sec:6_Conclusion}}

Following the completion of its neutrino-mass measurement campaigns, the KATRIN experiment will enter a new physics phase dedicated to the search for keV-scale sterile neutrinos with the TRISTAN detector. This transition introduces a fundamentally different measurement strategy based on high-rate and differential spectroscopy of the tritium $\upbeta$-decay spectrum over a broad energy range.

In this work, we have presented a sensitivity study of KATRIN in this forthcoming phase based on a dedicated forward-convolution framework that incorporates the dominant beamline, detector, and readout effects relevant for the measurement. For the baseline scenario of four months of detector runtime considered in this analysis, KATRIN is found to reach a statistical sensitivity at the level of $|U_{e4}|^2 \sim 10^{-6}$ for sterile-neutrino masses in the range of approximately (4--13)\,keV. 
When the major experimental systematic uncertainties are included, the sensitivity is reduced by up to a factor of 50, yielding a potential projected sensitivity at the level of $|U_{e4}|^2 \sim 2\times10^{-5}$ over the same mass range. 
Importantly, no single systematic effect dominates the sensitivity loss. Instead, the expected sensitivity emerges from the cumulative impact of several sub-dominant effects.

Overall, this study demonstrates that KATRIN equipped with the TRISTAN detector has the potential to provide a powerful laboratory probe of keV-scale sterile neutrinos that is independent of cosmological and astrophysical model assumptions, extending well beyond existing laboratory constraints. 
As demonstrated in this work, mismodeling of dominant systematic contributions can significantly degrade the sensitivity. Achieving the projected sensitivity will therefore require stringent control and validation of the systematic effect models. Nevertheless, this study provides a quantitative baseline for the TRISTAN phase and a clear prioritization of the systematic contributions that should be addressed. At the same time, the combination of ultra-high-statistics measurements with flexible beamline operation places the experiment in a uniquely strong position to constrain, and mitigate such effects.
The analysis framework established here provides a solid foundation for future modeling refinements as commissioning data become available.

 \section*{Acknowledgments \label{sec:7_Acknowledgements}}

We acknowledge the support of Helmholtz Association (HGF), Ministry for Education and Research BMBF (05A23PMA, 05A23PX2, 05A23VK2, and 05A23WO6), the doctoral school KSETA at KIT, Helmholtz Initiative and Networking Fund (grant agreement W2/W3-118), Max Planck Gesellschaft, and Deutsche Forschungsgemeinschaft DFG (GRK 2149 and SFB-1258 and under Germany's Excellence Strategy EXC 2094 – 390783311) in Germany; Ministry of Education, Youth and Sport (CANAM-LM2015056) in the Czech Republic; Istituto Nazionale di Fisica Nucleare (INFN) in Italy; Suranaree University of Technology (SUT), Thailand Science Research and Innovation (TSRI), the National Science, Research and Innovation Fund (NSRF) (Project No.\ 204211), and the National Science, Research and Innovation Fund via the Program Management Unit for Human Resources \& Institutional Development, Research and Innovation (grant B39G680010) in Thailand; and the Department of Energy through grants DE-FG02-97ER41020, DE-FG02-94ER40818, DE-SC0004036, DE-FG02-97ER41033, DE-FG02-97ER41041,  {DE-SC0011091 and DE-SC0019304 and the Federal Prime Agreement DE-AC02-05CH11231} in the United States. This project has received funding from the European Research Council (ERC) under the European Union Horizon 2020 research and innovation programme (grant agreement No. 852845). We thank the computing support at the Institute for Astroparticle Physics at Karlsruhe Institute of Technology, Max Planck Computing and Data Facility (MPCDF), and the National Energy Research Scientific Computing Center (NERSC) at Lawrence Berkeley National Laboratory.

\bibliography{8-Bibliography} 

\end{document}